\documentclass[twocolumn, tighten]{aastex631}
\pdfoutput=1 \usepackage{amsmath}
\usepackage{amsfonts}
\usepackage{amssymb}
\usepackage{mathtools}
\usepackage{graphics}
\usepackage{tabularx}
\usepackage{booktabs}
\usepackage{multirow}
\usepackage{txfonts}
\usepackage{graphicx}
 \usepackage{url}
\usepackage{soul}
\usepackage{bm}
\usepackage{xcolor}
\usepackage[utf8]{inputenc}
\usepackage[normalem]{ulem}
\usepackage{hyperref}

\newenvironment{system}%
{\left\lbrace\begin{array}{@{}l@{}}}%
{\end{array}\right.}

\shorttitle{Neutrino Emission from Luminous Fast Blue Optical Transients}
\shortauthors{Guarini, Tamborra \& Margutti}

\graphicspath{{./}{figures/}}

\begin{document}

\title{Neutrino Emission from Luminous Fast Blue Optical Transients}

\author[0000-0002-3744-8592]{Ersilia Guarini}
\affiliation{Niels Bohr International Academy \& DARK, Niels Bohr Institute, University of Copenhagen, Blegdamsvej 17, 2100, Copenhagen, Denmark}

\author[0000-0001-7449-104X]{Irene Tamborra}
\affiliation{Niels Bohr International Academy \& DARK, Niels Bohr Institute, University of Copenhagen, Blegdamsvej 17, 2100, Copenhagen, Denmark}

\author[0000-0003-4768-7586]{Raffaella Margutti}
\affiliation{Department of Astronomy, University of California, 501 Campbell Hall, Berkeley, CA 94720, USA}
\affiliation{Department of Physics, University of California, 366 Physics North MC 7300,
Berkeley, CA 94720, USA}

\begin{abstract}
Mounting evidence suggests that Luminous Fast Blue Optical Transients (LFBOTs) are powered by a compact object,  launching an asymmetric and fast outflow  responsible for the  radiation observed in the ultraviolet, optical, infrared, radio, and X-ray bands. 
Proposed scenarios aiming to explain the electromagnetic emission include an inflated  cocoon, surrounding a jet  choked in the extended stellar envelope. In alternative, the observed radiation  may arise from the disk formed by the delayed merger of a black hole with a Wolf-Rayet star. We explore the neutrino production in these scenarios, i.e.~internal shocks in a choked jet and  interaction between the outflow and the circumstellar medium (CSM). If observed on-axis, the choked jet provides the dominant contribution to the neutrino fluence.
Intriguingly,  the IceCube upper limit on the neutrino emission inferred from the closest LFBOT, AT2018cow,  excludes a region of the parameter space otherwise allowed by electromagnetic observations. After correcting for the Eddington bias on the observation of cosmic neutrinos, we conclude that the emission from an on-axis choked jet and CSM interaction is compatible with the detection of  two track-like neutrino events observed by the IceCube Neutrino Observatory  in coincidence with AT2018cow, and otherwise considered to be of atmospheric origin. While the neutrino emission from LFBOTs does not constitute the bulk of the  diffuse background of neutrinos observed by IceCube,  detection prospects of nearby LFBOTs with   IceCube   and the upcoming  IceCube-Gen2 are encouraging.
Follow-up neutrino searches will be crucial for unravelling the mechanism powering this emergent transient class.
\end{abstract}


\section{Introduction} \label{sec:intro}

The advent of time-domain astronomy  has led to the discovery of intriguing new classes of transients that evolve on time-scales $\lesssim 10$~days  \citep[e.g.][]{2010Sci...327...58P, Inserra:2019ciq, Modjaz:2019flw, Drout:2014dma, Arcavi:2015zie}.  Among these, Fast Blue Optical Transients (FBOTs)  \citep{Drout:2014dma, Arcavi:2015zie,Tanaka:2016ncv, DES:2018whm,Ho:2021fyb} exhibit  unusual features. They have a rise time of a few days in the optical---$t_{\rm{rise}}$ up to $3$~days, i.e.~much faster than typical supernovae~(SNe; e.g.~\citealt{Vallely:2020eub, Arcavi:2015zie, Ho:2021fyb})---and their spectrum remains blue and  hot throughout the whole evolution.

We focus on the subclass of optically luminous FBOTs (hereafter denoted with LFBOTs), 
with absolute peak magnitude $M_{\rm{peak}} < -20$~\citep{Ho:2020hwf, Coppejans:2020nxp, Ho:2021fyb}.  LFBOTs  have a rate in the local Universe $\lesssim 300$~Gpc$^{-3}$~yr$^{-1}$, i.e. $\lesssim 0.4 - 0.6 \%$ of core-collapse SNe (\citealt{Ho:2020hwf, Coppejans:2020nxp, Ho:2021fyb}). 
To date,  radio emission has been detected for five FBOTs, all belonging to the LFBOTs category: CSS161010, AT2018cow,  AT2018lug, AT2020xnd, and AT2020mrf. LFBOTs have been detected in the hard X-ray band as well, though not yet in gamma-rays (i.e.~with energies $> 200$ keV)~\citep{Ho:2018emo, Margutti:2018rri, Coppejans:2020nxp,  Ho:2020hwf, Yao:2021qqz, Bright:2021lod}.

The radio signal associated with LFBOTs is consistent with synchrotron radiation in the self-absorption regime, arising from the forward shock developing when the ejecta interact with  the circumstellar medium (CSM). Broad hydrogen (H) emission features have been observed in the spectra of some LFBOTs, i.e. AT2018cow~\citep{Perley:2018oky, Margutti:2018rri} and CSS161010~\citep{Coppejans:2020nxp}. Moreover, combined observations in the optical and radio bands suggest that the fastest component of the outflow is moving with  speed $0.1 c \lesssim v_{\rm{f}} \lesssim 0.6 c$~\citep{Perley:2018oky, Ho:2018emo, Margutti:2018rri,  Yao:2021qqz, Bright:2021lod}.

As for X-rays, the spectrum exhibits a temporal evolution and a high variability that is challenging to explain by invoking   external shock interaction. Rather, the X-ray emission might be powered by a rapidly evolving compact object (CO), like a magnetar or a black hole, or a deeply embedded shock~\citep{Ho:2018emo, Margutti:2018rri}. In addition, interaction with the CSM cannot simultaneously explain  the ultraviolet, optical, and infrared spectral features, e.g.~the rapid rise of the light curve and its luminosity ($L_{\rm{opt}} \simeq 10^{44}$~erg s$^{-1}$), as well as the receding photosphere  observed for AT2018cow at late times~\citep{Perley:2018oky} and typically associated with an increase of the effective temperature. In the light of this growing set of puzzling data, multiple  sites  might be at the origin of the observed electromagnetic emission across  different wavebands, together with an asymmetric outflow embedding the CO~\citep{Margutti:2018rri}. An additional piece of evidence of the presence of a CO might be the persistent ultraviolet source observed at the location of AT2018cow~\citep{Sun:2022vln}. The presence of a CO may also be supported by the observation in AT2018cow of high-amplitude quasi periodic oscillations in soft X-rays~\citep{Pasham:2021rec}.

Several interpretations   of   LFBOT observations have been proposed,  such as shock interaction of an outflow with  dense CSM (e.g.~\citealt{Fox:2019zmd, Pellegrino:2021azo, Leung:2020msx, Xiang:2021fal}\footnote{We note, however, that the broad-band X-ray spectrum of AT2018cow is unlike the thermal spectra of interacting SNe, and shows instead clear non-thermal features.}); reprocessing  of X-rays emitted from a central engine within a polar outflow (e.g.~\citealt{Margutti:2018rri, Chen:2022ihf, 2021MNRAS.507.1092C, Piro:2020gqr, Uno:2020vtl, Liu:2018hft, Perley:2018oky, Kuin:2018avi}); a neutron star engulfed in the extended envelope of a massive red supergiant, leading to  common envelope evolution and formation of a jetted SN~\citep{Soker:2018msh} or a related  impostor~\citep{Soker:2022gna}; emission from the accretion disk originating from the collapse of a massive star into a black hole~\citep{Kashiyama:2015qfa, Quataert:2018gnt}  or from the electron-capture collapse to a neutron star following the merger of a  ONeMg white dwarf with another  white dwarf~\citep{Lyutikov:2018pkv,Lyutikov:2022xri}.  Each of the aforementioned scenarios may only reproduce some of the observed features of LFBOTs.

Recently, two  models have been  proposed  in the attempt of explaining the multi-wavelength emission of AT2018cow. \cite{Gottlieb:2022old} invoke the collapse of a massive star, possibly not completely H-stripped, which launches a jet. The jet may be off-axis or  choked in the extended stellar envelope and, therefore, not directly visible; to date, direct associations between jets and LFBOTs are  lacking and constraints have been set for AT2018cow~\citep{Bietenholz:2019ptf}. The jet interacts with the stellar envelope, inflating the cocoon  surrounding the jet; the cocoon expands, breaks out of the star and cools, emitting  in the ultraviolet, optical, and infrared. \cite{Metzger:2022xep} considers a delayed Wolf-Rayet star--black hole merger following a failed common envelope phase. This leads to the formation of an asymmetric CSM,  dense in the equatorial region, and less dense in the polar one. The scenarios proposed by \cite{Gottlieb:2022old} and \cite{Metzger:2022xep}  successfully fit the ultraviolet, optical and infrared spectra of AT2018cow; \cite{Metzger:2022xep} also provides a fit to the X-ray data of AT2018cow. However, it is yet to be quantitatively proven that the off-axis jet scenario of~\cite{Gottlieb:2022old} is consistent with radio observations; no fit to the radio data is provided in~\cite{Metzger:2022xep}. 
It is unclear whether these models could explain the late time hot and luminous ultraviolet emission ($L_{\rm{UV}} \gtrsim 2.7 \times 10^{34}$~erg s$^{-1}$) detected in the proximity of AT2018cow~\citep{Sun:2022vln}. \cite{Metzger:2022xep} provides a possible explanation to this persistent emission as the late time radiation from the accretion disk surrounding the black hole resulting from the Wolf-Rayet star--black hole merger. Further observations in the direction of AT2018cow will eventually confirm this conjecture.

In order to unravel the nature of the engine powering LFBOTs, a multi-messenger approach may provide a fresh perspective. In particular, the neutrino signal could carry signatures of the  mechanisms powering LFBOTs.
Since the first detection of   high-energy neutrinos of astrophysical origin by the IceCube Neutrino Observatory, follow-up searches are ongoing to pinpoint the electromagnetic counterparts associated to the IceCube neutrino events~\citep{IceCube:2020mzw,Fermi-LAT:2019hte,Acciari:2021YA,VERITAS:2021mjg,Necker:2022tae,Stein:2022rvc}. A dozen of neutrino events have been associated in likely coincidence with blazars, tidal distruption events or superluminous supernovae~\citep{IceCube:2018dnn,Giommi:2020viy,Franckowiak:2020qrq,Fermi-LAT:2019hte,Krauss:2018tpa,Kadler:2016ygj,Stein:2020xhk,Reusch:2021ztx,Pitik:2021dyf}.
As for LFBOTs,  the IceCube Neutrino Observatory reported the detection of two track-like muon neutrino events in spatial coincidence with AT2018cow  in the $3.5$~days following the optical detection. These neutrino events  could be statistically compatible with the expected number of atmospheric neutrinos--$0.17$ events~\citep{2018ATel11785....1B}.

As the number of  LFBOTs detected electromagnetically increases,  the related neutrino emission remains poorly explored. 
\cite{Fang:2018hjp} pointed out that, if  AT2018cow is powered by a magnetar,  particles accelerated in the magnetar wind may escape the ejecta at ultrahigh energies.
Within the models proposed in~\cite{Gottlieb:2022old,Metzger:2022xep}, additional sites should be taken into account  for what concerns neutrino production.
For example, if a choked jet powered by the central CO is harbored within the LFBOT~\citep{Gottlieb:2022old},  we would not observe any prompt gamma-ray signal. Nevertheless, efficient proton acceleration could take place leading to the production of  TeV--PeV neutrinos~\citep{Murase:2013ffa, He:2018lwb, Meszaros:2001ms, Razzaque:2004yv, Ando:2005xi, Nakar:2015tma, Senno:2015tsn, Xiao:2014vga,Fasano:2021bwq,Tamborra:2015fzv,Denton:2017jwk}. 
In addition,  \cite{Gottlieb:2022old,Metzger:2022xep} predict  fast ejecta propagating in the CSM with velocity $v_{\rm{f}} \gtrsim 0.1 c$. Protons may be accelerated at the shocks between the ejecta and the CSM leading to  neutrino production, similar to  what foreseen for SNe~\citep{Murase:2010cu, Pitik:2021dyf, Petropoulou:2017ymv, Petropoulou:2016zar, 2011arXiv1106.1898K, Murase:2013kda, Cardillo:2015zda,Zirakashvili:2015mua, Murase:2020lnu,Sarmah:2022vra} or trans-relativistic SNe~\citep{Kashiyama:2013qet, Zhang:2018agl}, probably powered by a choked jet as it may be the case for LFBOTs. 
Neutrinos produced from LFBOTs could be detectable by  the IceCube Neutrino Observatory and the upcoming  IceCube-Gen2, aiding to pin down the mechanisms powering LFBOTs~\citep{Murase:2019tjj,Fang:2020bkm}. 

Our work is organized as follows. In Sec.~\ref{sec:twomodels}, we discuss the most promising particle acceleration sites for the models proposed in \cite{Gottlieb:2022old} and \cite{Metzger:2022xep} (a choked jet and/or a fast outflow emitted by the CO that propagates outwards in the CSM).
Section~\ref{sec:parameters} summarizes the model parameters inferred for AT2018cow and CSS161010 from electromagnetic observations. Section~\ref{sec:neutrinoProduction} focuses on the production of high-energy neutrinos.
In Section~\ref{sec:results}, we present our findings for the neutrino  signal expected at Earth from AT2018cow and CSS161010 and discuss the corresponding detection prospects. The contribution of LFBOTs to the neutrino diffuse background is presented in Sec.~\ref{sec:diffuse}. Finally, we conclude in Sec.~\ref{sec:conclusions}. The most relevant proton and meson cooling times are outlined in Appendix~\ref{sec:A}.

\section{Particle acceleration sites}\label{sec:twomodels}
In this section, we outline the mechanisms proposed in \cite{Gottlieb:2022old} (hereafter named ``cocoon model'') and \cite{Metzger:2022xep} (hereafter ``merger model'') for powering LFBOTs that could also host sites of particle acceleration. First, we consider  a jet launched by the central engine and  choked in the  extended stellar envelope. Then, we focus on the interaction  between the fast ejecta  and the CSM.

\subsection{Choked jet}\label{sec:chokedjet}
\begin{figure*}
\includegraphics[width=0.48\textwidth]{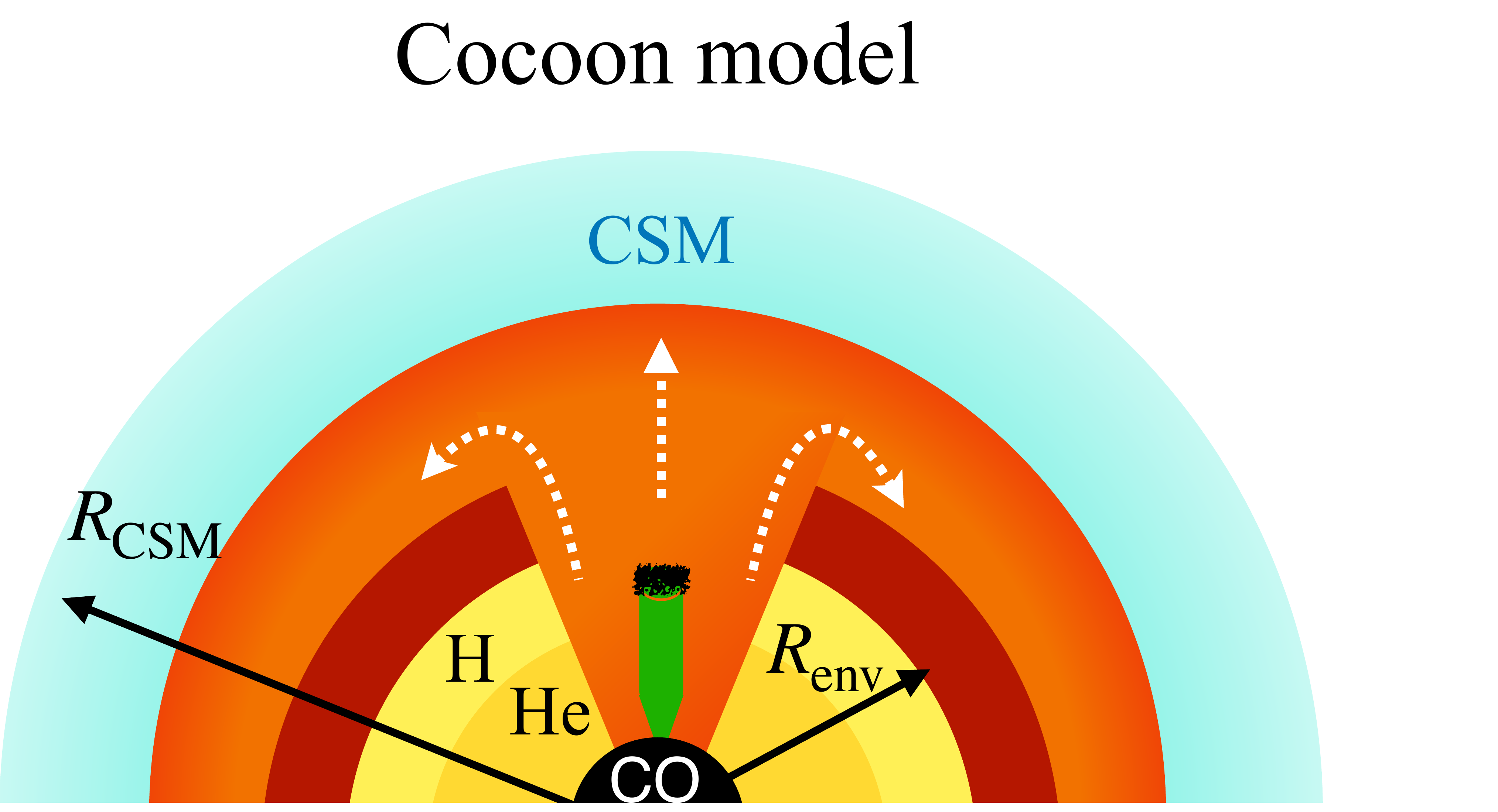}
\hspace*{0.7cm}\includegraphics[width=0.50\textwidth]{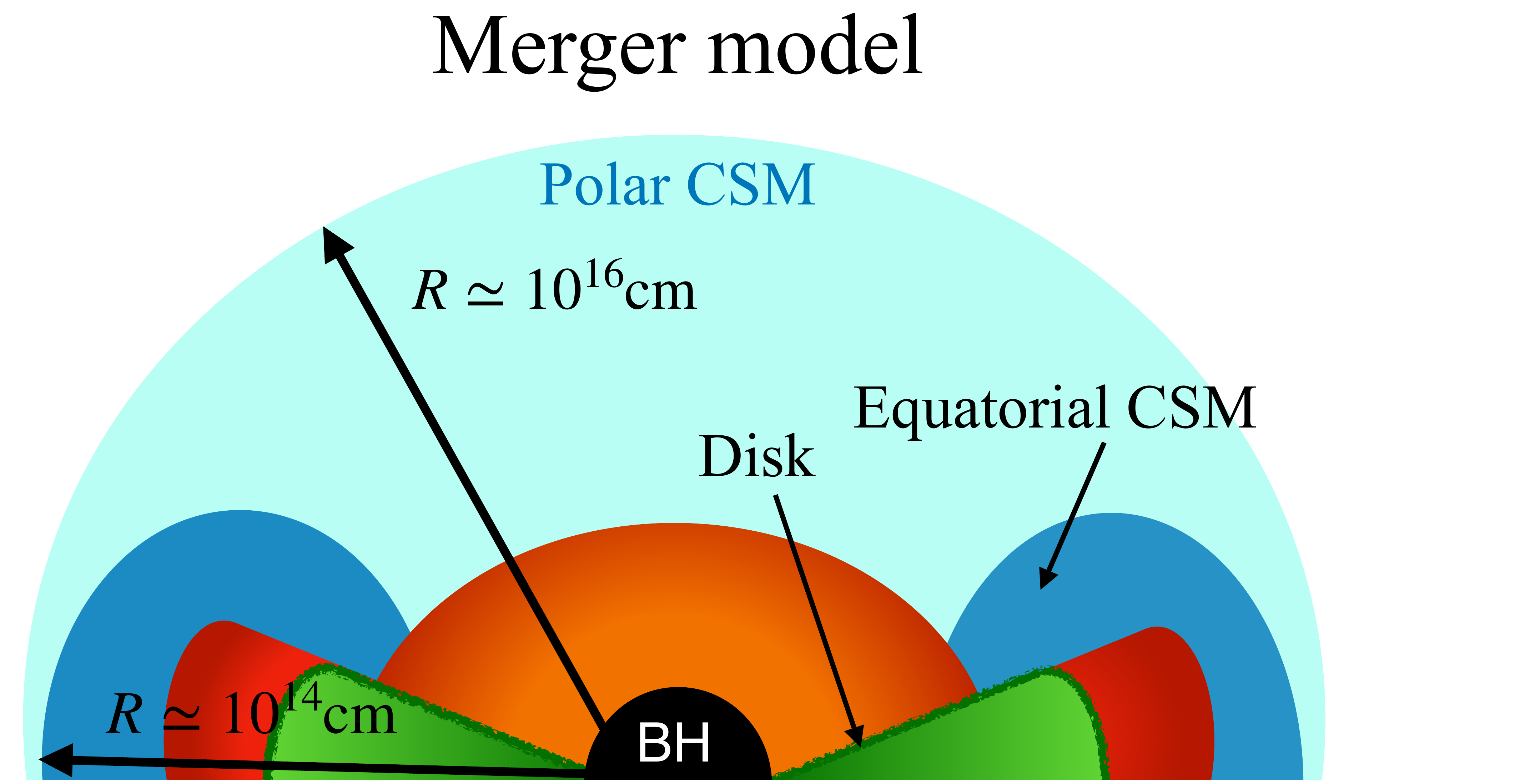}
\hspace*{-0.03cm}\includegraphics[width=0.45\textwidth]{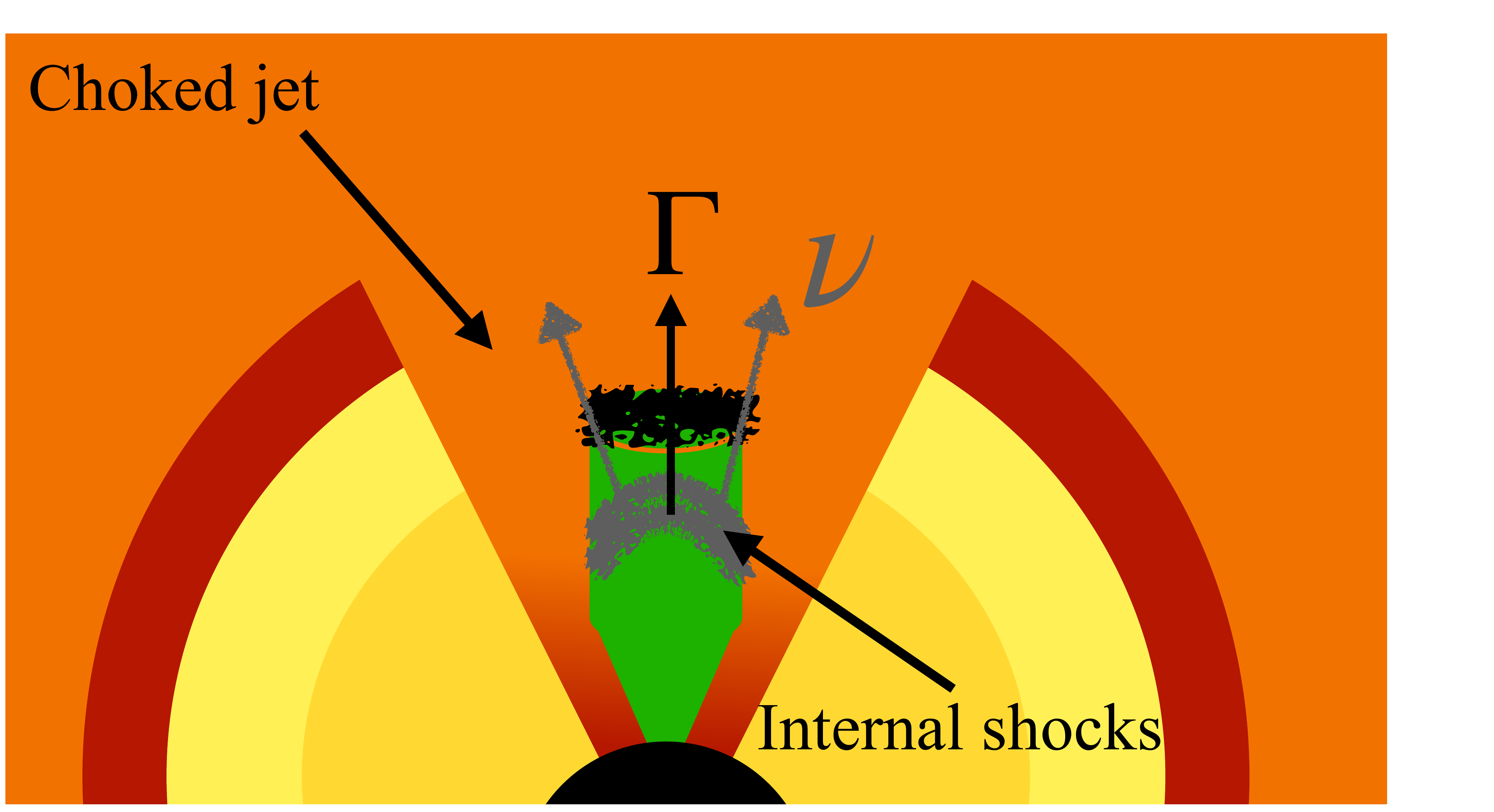}
\hspace*{1.1cm}\includegraphics[width=0.49\textwidth]{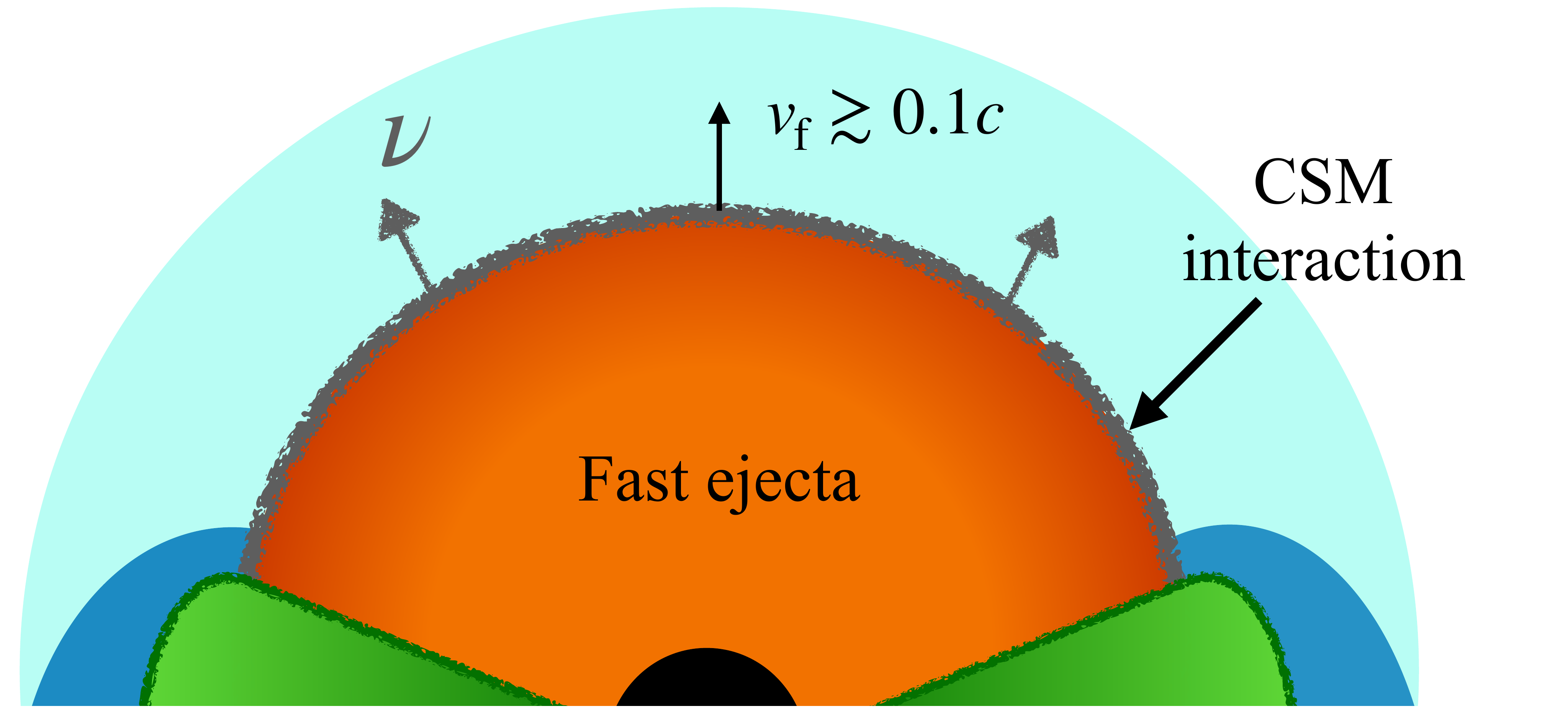}
\hspace*{-0.1cm}\includegraphics[width=0.49\textwidth]{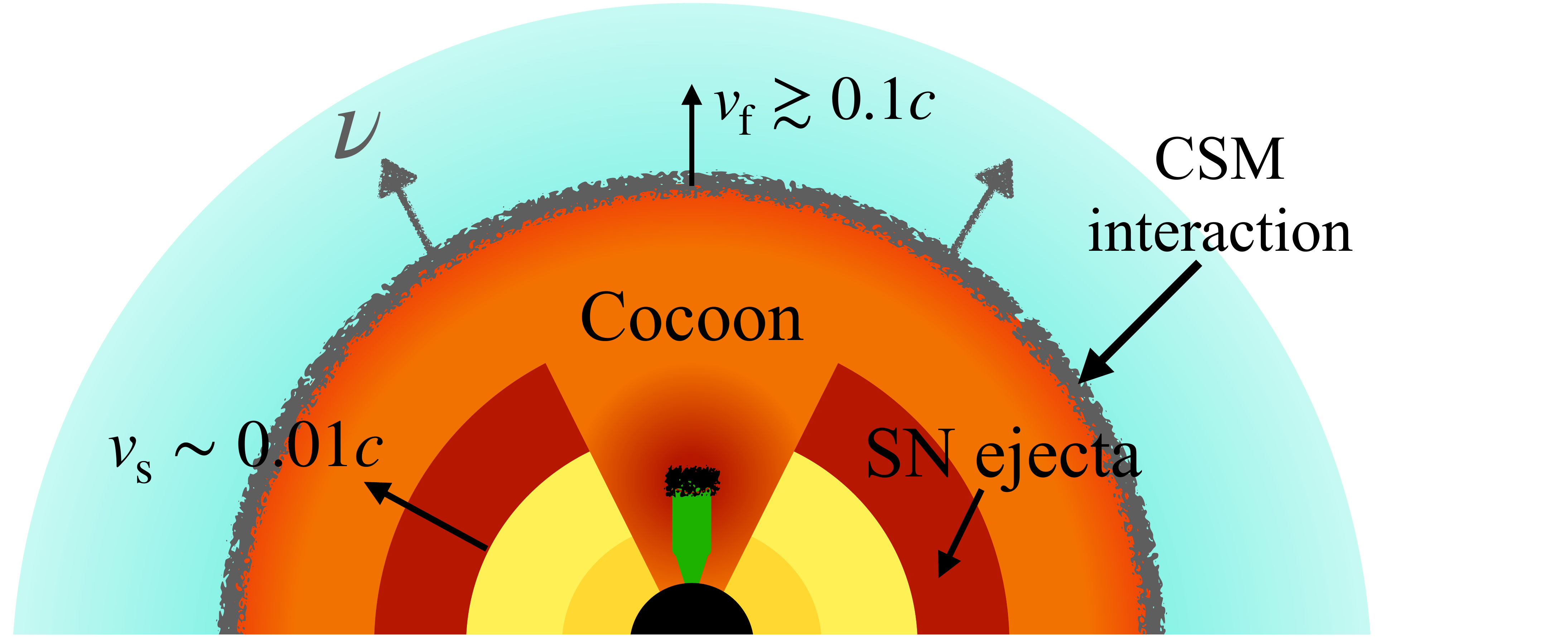}
\hspace*{1.4cm}\includegraphics[width=0.44\textwidth]{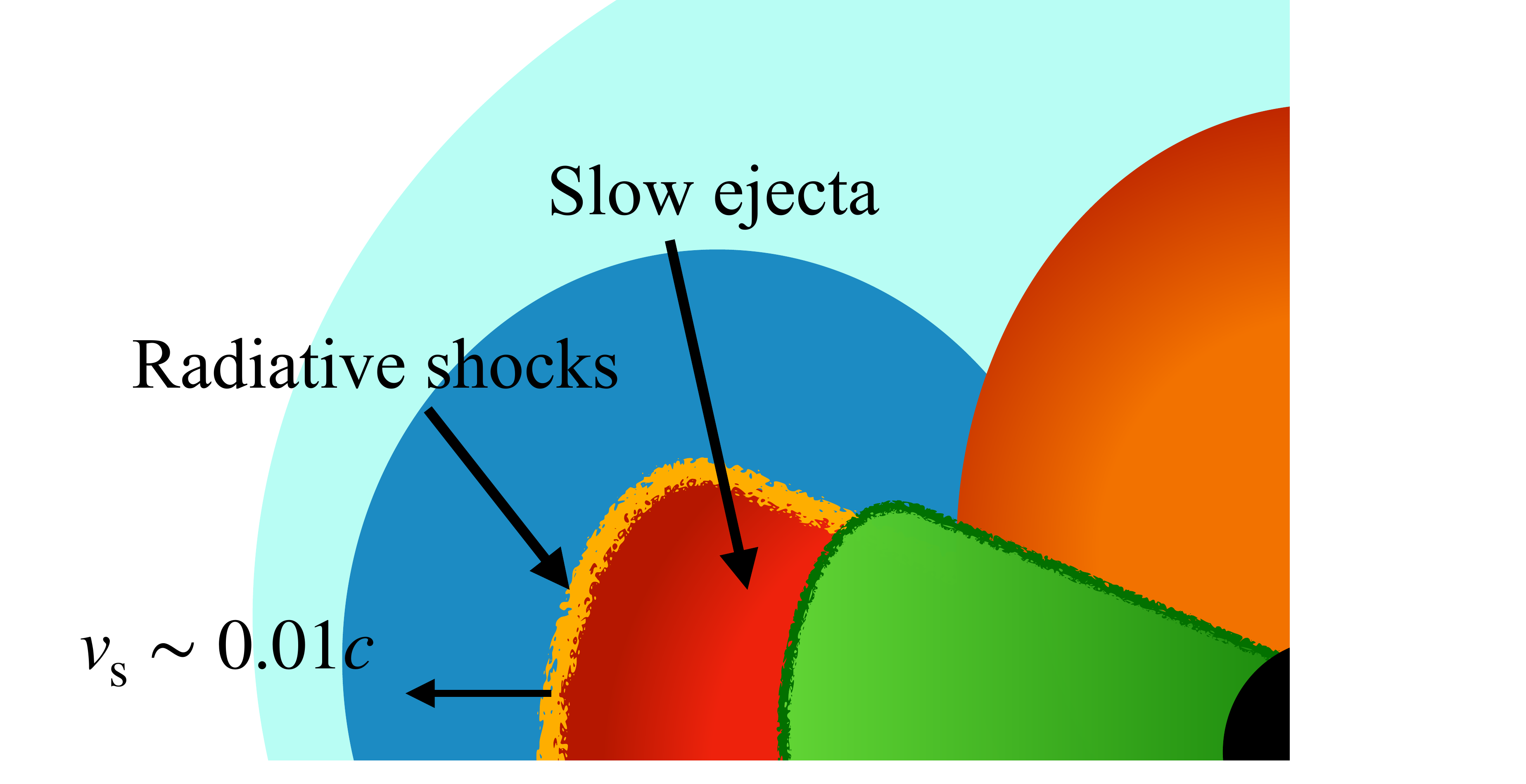}
\caption{Cartoons of the cocoon model (left panels, \cite{Gottlieb:2022old}) and merger model (right panels, \cite{Metzger:2022xep}), not to scale. For the sake of simplicity, we  show only the upper half section of the FBOT. \textit{Top left panel}: A massive star collapses, forming a CO (black region). The CO is surrounded by helium (He) and H envelopes (regions with yellow hues). The progenitor core ($R_\star \sim 10^{11}$~cm) is surrounded by an extended envelope (of radius $R_{\rm{env}}$).  \textit{Middle left panel}: The jet (green) is launched near the surface of the CO and it is choked in the extended envelope. Internal shocks occur in the proximity of the jet head (gray), where neutrinos can be produced. \textit{Bottom left panel}: The jet inflates the cocoon (orange region); the latter breaks out from the stellar surface and interacts with the CSM (aqua outer region). The fastest component of the cocoon moves with $v_{\rm{f}} \gtrsim 0.1 c$, while its slow component (red region; SN ejecta) propagates with $v_{\rm{s}} \simeq 0.01 c$ in the equatorial direction. While the fast component of the cocoon propagates into the CSM, collisionless shocks take place (gray line surrounding the cocoon); here, neutrinos may be produced. Even though the geometry of the cocoon is not perfectly spherical, we assume   spherical symmetry for the sake of simplicity in the analytical treatment of the problem; see main text.  \textit{Top right panel}: As a result of  the Wolf-Rayet star-black hole merger, a black hole forms (BH; black), surrounded by an accretion disk (green region). The equatorial dense CSM (blue region) extends up to $\simeq 10^{14}$~cm, while the polar (aqua region) CSM extends up to $\simeq 10^{16}$~cm. \textit{Middle right panel}:  The disk emits a fast outflow (orange region) propagating in the polar direction with $v_{\rm{f}} \simeq 0.1 c$  into the CSM. Here, collisionless shocks (gray line) occur and neutrino production takes place. \textit{Bottom right panel}:  The slow outflow (red shell) is emitted from the disk in the equatorial direction, and it propagates with $v_{\rm{s}} \simeq 0.01 c$ into the dense equatorial CSM. Here, radiative shocks take place (orange line) and neutrino production is negligible with respect to the one from the polar outflow.}
\label{fig:cartoon}
\end{figure*}
\cite{Gottlieb:2022old} propose that LFBOTs arise from the collapse of massive stars that result in the formation of a central CO, possibly harboring a relativistic jet, as shown in the left panels  of Fig.~\ref{fig:cartoon}.  If the jet were to successfully drill through the  stellar envelope, it would  break out and give rise to a gamma-ray bright signal. Nevertheless, no prompt emission has been detected in association with LFBOTs, suggesting that a successful jet could be disfavored~\citep{Margutti:2018rri, Coppejans:2020nxp}. The non detection of  gamma-rays hints that an extended envelope, probably not fully  H-depleted in order to explain the broad emission features observed in some LFBOT spectra [AT2018cow~\citep{Perley:2018oky, Margutti:2018rri} and CSS161010~\citep{Coppejans:2020nxp}], may engulf the stellar core, extending up to $R_\star \simeq 10^{11}$~cm~\citep{Gottlieb:2022old}. In this case, the jet could be  choked, as displayed in the middle left panel of Fig.~\ref{fig:cartoon}. 

We consider a collapsing star that has not  lost its  H envelope completely  and it is surrounded by an extended shell of radius $R_{\rm{env}} \simeq 3 \times 10^{13}$~cm and mass $M_{\rm{env}} \simeq 10^{-2} M_\odot$~\citep{Senno:2015tsn}. The modeling of the extended H envelope mass is inspired by  partially stripped SNe~\citep[e.g.][]{Gilkis:2021uht, Nakar:2015tma, Sobacchi:2017wcq}. We fix the value of $M_{\rm{env}}$ to avoid to deal with several free parameters  (see Sec.~\ref{sec:parameters}) and  leave to  future work the assessment  of the dependence of the neutrino signal  on the mass of the extended envelope. For the extended envelope we consider the following density profile~\citep{Nakar:2015tma}:
\begin{equation}
      \rho_{\rm{env}}(R) = \rho_{\rm{env}, 0} \biggl( \frac{R}{R_{\rm{env}}} \biggl)^{-2} \; ,
\end{equation}
where $\rho_{ \rm{env}, 0} = M_{\rm{env}} \left[ \int^{R_{\rm{env}}} dR 4 \pi R^2 \rho_{\rm{env}}(R) \right] ^{-1}$ and $R$ is the distance from the CO. We assume a fixed density profile for the extended envelope due to the lack of knowledge on its features; further investigations on the impact of this assumption on the neutrino signal is left to  future work. Nevertheless, we expect that neutrino telescopes will not be sensitive to this dependence, see e.g.~\cite{Xiao:2014vga}.
The jet is launched near the surface of the CO~\footnote{We rely on three different reference frames throughout this paper: the CO frame, the observer frame and the jet comoving frame. In order to distinguish among them, each quantity in each of these frames is denoted as $\tilde{X}, X, X^\prime$, respectively.}, with luminosity $\tilde{L}_j$, narrow opening angle $\theta_j$. 

For fixed $\theta_j$, the dynamics of the jet only depends  on the isotropic equivalent quantities. Hence,  it is convenient to define the isotropic equivalent luminosity of the jet: $\tilde{L}_{j}^{\rm{iso}} = \tilde{L}_j/(\theta_j^2/4)$. Note that the isotropic equivalent quantities are always defined in the CO frame;  for the sake of clarity we  keep the twiddle notation throughout the paper.

While the jet pierces through the stellar envelope, two shocks develop: a reverse shock, propagating back to the core of the jet, and a forward shock, propagating into the external envelope. The region between the two shocks constitutes the jet head. Denoting with $\Gamma$ the Lorentz factor of the un-shocked jet plasma (i.e., the bulk Lorentz factor of the jet) and with $\Gamma_h$ the one of the jet head, the relative Lorentz factor  is~\citep{He:2018lwb}:
\begin{equation}
{\Gamma}_{\rm{rel}}= \Gamma \Gamma_h (1- \beta \beta_h) \; ,
\end{equation}
where $\beta = \sqrt{1-1/ \Gamma^2}$ and $\beta_h = \sqrt{1-1/ \Gamma_h^2}$.   
For a non relativistic jet head: $\Gamma_h \simeq 1$, which implies ${\Gamma}_{\rm{rel}} \simeq \Gamma$; this assumption is valid for the region of the parameter space of interest, as discussed in Sec.~\ref{sec:parameters}.

From the shock jump conditions, the energy density in the shocked envelope region and in the shocked jet plasma at the position of the jet head $\tilde{R}_h \equiv R_h$,  respectively, are~\citep{1976PhFl...19.1130B, Sari:1995nm}: 
\begin{eqnarray}
e_{\rm{sh, env}} &=& (4 \Gamma_h + 3) (\Gamma_h - 1) \rho_{\rm{env}} (R_h) c^2 \ , \label{eq:energy_env} \\
e_{\rm{sh, j}} &= &(4 {\Gamma}_{\rm{rel}} + 3) ({\Gamma}_{\rm{rel}} -1) n^\prime_j (R_h) m_p c^2\ . \label{eq:energy_shj}  
\end{eqnarray} 
Here $n^\prime_j =  \tilde{L}_{j}^{\rm{iso}}/(4 \pi R^2 m_p c^3 \Gamma^2)$ is the comoving particle density of the un-shocked jet. 
Equating $e_{\rm{sh, ext}} = e_{\rm{sh, j}}$ and expanding around $\Gamma_h$ for the non-relativistic case, we obtain the speed of the jet head:
\begin{equation}
{v}_h \simeq \left[ \frac{ \tilde{L}_{j}^{\rm{iso}}}{(4 \Gamma_h +3) \pi c \rho_{\rm{env}}(R_h) R_h^2} \right]^{1/2} \; .
\label{eq:head_speed}
\end{equation}
Since the jet head is non relativistic, its position at the time ${t}$ is ${R}_h \simeq {v}_h t /(1+z) = {v}_h \tilde{t}$, where $z$ is the redshift of the source~\footnote{In the literature a geometrical correction factor of $2$ is often  considered in the relations between the radius of the head and the time, see e.g.~\cite{He:2018lwb}; nevertheless, this does not affect our findings.}. Plugging the last expression in Eq.~\ref{eq:head_speed} we obtain the position of the jet head at the end of the jet lifetime $\tilde{t}_j$,
\begin{equation}
{R}_h \simeq   \left[   \frac{ \tilde{t}_j^2  \tilde{L}_{j}^{\rm{iso}}}{(4 \Gamma_h +3)  \pi c \rho_{ \rm{env}, 0} R_{\rm{env}}^{2}} \right]^{1/2} \; .
\label{eq:head_radius}
\end{equation}
If $R_h < R_{\rm{ext}}$, the jet is choked inside the stellar envelope. 

The jet consists of several shells moving with different velocities. This implies that internal shocks may take place in the jet at  $R_{\rm{IS}} \lesssim R_h $, when a fast shell catches up and merges with a slow shell. If $\Gamma_{\rm{r}} \simeq \left( \Gamma_{\rm{fast}}/ \Gamma_{\rm{merg}} + \Gamma_{\rm{merg}}/ \Gamma_{\rm{fast}} \right) /2 $ is the relative Lorentz factor between the fast (moving with $\Gamma_{\rm{fast}}$) and the merged shell (moving with $\Gamma_{\rm{merg}}$) in the jet, efficient particle acceleration at the internal shock takes place only if~\citep{Murase:2013ffa}
\begin{equation}
n^{\prime}_{p} \sigma_T R_{\rm{IS}}/ \Gamma \lesssim \rm{min} \bigl[\Gamma_{\rm{r}}^2, 0.1 C^{-1} \Gamma_{\rm{r}}^3 \bigr]\ ,
\label{eq:rad_condition}
\end{equation}
where $C = 1 + 2 \rm{ln}{\Gamma^2_{\rm{r}}}$ is a constant taking into account pair production and $n^\prime_{p} \simeq n^\prime_j$ is the proton density of the un-shocked jet material. If Eq.~\ref{eq:rad_condition} is not satisfied,  the internal shock is radiative and particle acceleration is not efficient~\citep{Murase:2013ffa}.  We assume that internal shocks approach the jet head, i.e. $R_{\rm{IS}} \simeq R_h$~\citep{He:2018lwb}.

\subsubsection{Photon energy distribution}
Electrons can be accelerated at the reverse shock between the shocked and the un-schocked jet plasma. Then, they heat up and rapidly thermalize due to the high Thomson optical depth of the jet head
\begin{equation}
\tau_{T, h} = n_{e, \rm{sh, j}} \sigma_T \frac{R_h}{\Gamma_h} \gg 1 \; , 
\label{eq:optical_depth}
\end{equation}
where $n_{e, \rm{sh, j}} = (4 \Gamma_h +3) n^\prime_j $ is the electron number density of the shocked jet plasma. Therefore, the electrons in the jet head lose all their energy ($\epsilon_e^{\rm{RS}} e_{\rm{sh}, j}$) through thermal radiation, with $e_{\rm{sh}, j}$ defined as in Eq.~\ref{eq:energy_shj} and $\epsilon_e^{\rm{RS}}$ being the fraction of the energy density that goes into the electrons accelerated at the reverse shock. The temperature of the emitted thermal radiation, in the jet head comoving frame, is~\citep{Razzaque:2005bh, Tamborra:2015fzv}
\begin{equation}
k_B T_h \simeq \biggl( \frac{30 \hbar^3 c^2 \epsilon_e^{\rm{RS}} \tilde{L}_j^{\rm{iso}}}{4 \pi^4 R_h^2} \biggr)^{1/4} \; ,
\label{eq:headT}
\end{equation}
with $k_B$ being the Boltzmann constant. 
Thus, the head appears as a blackbody  emitting at temperature $k_B T^{\prime}_{\rm{IS}} = {\Gamma}_{\rm{rel}} k_B T_h$ in the comoving frame of the un-shocked jet. The density of thermal photons in the jet head is
\begin{equation}
n_{\gamma, h} = \frac{19 \pi}{(hc)^3} (k_B T_h)^3 \; . 
\label{eq:photonHead}
\end{equation}
As the internal shock approaches the head of the jet, a fraction $f_{\rm{esc}} = 1/ \tau_{T, h}$ of  thermal photons escapes in the internal shock~\citep{Murase:2013ffa}, where their number density  is boosted by  ${\Gamma}_{\rm{rel}}$: 
\begin{equation}
n^\prime_{\gamma, \rm{IS}} \simeq {\Gamma}_{\rm{rel}} f_{\rm{esc}} n_{\gamma, h} \; .
\end{equation}
The resulting energy distribution of thermal photons in the un-shocked jet comoving frame is [in units of GeV$^{-1}$ cm$^{-3}$]:
 \begin{equation}
    n^\prime_{\gamma}(E^\prime_\gamma) = \frac{{\rm{d}}^2 {N}_\gamma}{{\rm{d}} E^\prime_\gamma {\rm{d}} V^\prime} = A^\prime_{\gamma, j} \frac{E^{\prime -2}_{\gamma}}{ e^{E^\prime_\gamma /( k_B T^{\prime}_{\rm{IS}} )}-1} \; ,
    \label{eq:BBspectrum}
\end{equation}
where $A^\prime_{\gamma, j} = n^\prime_{\gamma, \rm{IS}} \left[ \int_0^{\infty} d E^\prime_\gamma n^\prime_{\gamma}(E^\prime_\gamma) \right]^{-1}$. 

\subsubsection{Proton energy distribution}
Protons are  accelerated to a power law distribution at the internal shock, even though the mechanism responsible for particle acceleration is still under debate (e.g.~\cite{Sironi:2013ri, Guo:2014via,  Nalewajko:2015gma,Petropoulou:2018bvv,Kilian:2020yyw}). The injected proton distribution in the jet comoving frame is [in units of GeV$^{-1}$ cm$^{-3}$] 
\begin{equation}
n^\prime_p(E^\prime_p) \equiv \frac{{\rm{d}}^{2}N^\prime_{\rm{p}}}{{\rm{d}}E^\prime_{\rm{p}}{\rm{d}}V^\prime} =  A^\prime_p E^{^\prime -k_p}_p \exp \left[-\left( \frac{E^\prime_p}{E^\prime_{p, \rm{max}}} \right)^{\alpha_p} \right] \Theta(E^\prime_p - E^\prime_{p, \rm{min}}) \ ,
\label{eq:protonJET}
\end{equation}
where $k_p$ is the proton spectral index, $\alpha_p = 1$  simulates an exponential cutoff~\citep{2001RPPh...64..429M}, and $\Theta$ is the Heaviside function. The value of $k_p$ is highly uncertain: it is estimated to be $k_p \simeq 2$ from non-relativistic shock diffusive acceleration theory~\citep{Matthews:2020lig}, while  it is expected to be $k_p \simeq 2.2$ from Monte Carlo simulations of ultra-relativistic shocks~\citep{Sironi:2013ri}. In this work, we assume $k_p \simeq 2$. 

The normalization constant is $A^\prime_p = \epsilon_p \epsilon_{\rm{d}} e^\prime_j \left[ \int_{E^\prime_{p, \rm{min}}}^{E^\prime_{p, \rm{max}}} dE^\prime_p  E^\prime_p n^\prime_p(E^\prime_p)\right] ^{-1}$, where $\epsilon_{\rm{d}}$ is the fraction of the comoving internal energy density of the jet $e^\prime_j =  \tilde{L}_{j}^{\rm{iso}} / (4 \pi R_{\rm{IS}}^2 c \Gamma^2)$ which is dissipated at the internal shock, while $\epsilon_p$ is the fraction of this energy that goes in accelerated protons. We rely on a one-zone model for  the emission from internal shocks and omit any radial evolution of the properties of the colliding shells. Hence, we assume that the dissipation efficiency $\epsilon_{\rm{d}}$ is constant~\citep[e.g.][]{Guetta:2000ye, Pitik:2021xhb}. Note, however, that $\epsilon_{\rm{d}}$  depends on the details of the collision, i.e.~the relative Lorentz factor between the colliding shells and their mass (see e.g.~\cite{Daigne:1998xc, Kobayashi:1997jk}).

The minimum energy of accelerated protons is $E^\prime_{p, \min} = m_p c^2$, while $E^\prime_{p, \rm{max}}$ is the maximum energy up to which protons can be accelerated at the internal shock. The latter is fixed by the condition that the proton acceleration timescale $t^{\prime -1}_{p, \rm{acc}}$ is smaller that the total cooling timescale $t^{\prime -1}_{p, \rm{cool}}$. For details on the cooling timescales of protons, see Appendix~\ref{sec:A}. At the internal shock, the fraction $\epsilon_B$ of the dissipated jet internal energy is given to the magnetic field: $B^{\prime}= \sqrt{8 \pi \epsilon_B \epsilon_{\rm{d}} e^\prime_{j}}$. 

\subsection{Interaction with the circumstellar medium}\label{sec:pp}
While the presence of a choked jet is  uncertain because of the lack of electromagnetic evidence~\citep{Bietenholz:2019ptf}, the existence of  fast ejecta launched by the central engine and moving with $v_{\rm{f}} \gtrsim 0.1 c$ is supported by observations  in the radio band~\footnote{It is worth noticing that the speed for the ejecta is very similar to the one of core-collapse SNe. Nevertheless, LFBOTs have been observed with fast ejecta speeds up to $v_{\rm{f}} \simeq 0.6 c$, see e.g.~\cite{Coppejans:2020nxp}. This feature makes these transients different from core-collapse SNe.}. 
The origin of the ejecta is still unclear and under debate. In the following, we discuss several viable mechanisms for the production of a fast outflow expanding outwards in the CSM.
\begin{itemize}
\item In the cocoon model presented in~\cite{Gottlieb:2022old} (see left panels of Fig.~\ref{fig:cartoon}),  
as the jet propagates in the stellar envelope (Sec.~\ref{sec:chokedjet}), a double-layered structure, the cocoon, forms around the jet, see e.g.~\cite{2011ApJ...740..100B}. 
The cocoon   breaks out from the star and expands in the surrounding CSM~\citep{Gottlieb:2021srg}. 
The interaction between the CSM and the cocoon is responsible for the observed radio signal. 
It is expected that the cocoon's ejecta are stratified in velocity, and the fastest component propagates with $v_{\rm{f}} \gtrsim 0.1 c$. Since we assume that the jet is choked in the extended stellar envelope and far from the stellar core, the fast component of the cocoon does not have any relativistic component moving with Lorentz factor $\Gamma_{\rm{f}} \sim 3$~\citep{Gottlieb:2021srg}. 
In addition to the fast ejecta, the outflow contains a slow component moving with $v_{\rm{s}} \lesssim 0.01 c$. This component might be the slow part of the SN ejecta accompanying the jet launching. Note that there might be a faster component of the SN ejecta, but the radio signal is probably dominated by the cocoon emission~\citep{Gottlieb:2022old}.

\item The merger model proposed in~\cite{Metzger:2022xep} (see right panels of Fig.~\ref{fig:cartoon}) invokes a Wolf-Rayet--black hole merger following a failed common envelope  phase. This leads to  a highly asymmetric CSM: a very dense region extends  up to $R\simeq 10^{14}$~cm around the equator and a less dense component extends up to $R \simeq 10^{16}$~cm in the polar direction. The asymmetric CSM is clearly required by electromagnetic observations of AT2018cow~\citep{Margutti:2018rri} and the energetics  of the fastest ejecta of CSS161010~\citep{Coppejans:2020nxp}. An accretion disk forms as a result of the  merger;  slow ejecta  in the equatorial direction move with $v_{\rm{s}} \simeq 0.01 c$, and the fast component  in the polar plane has $v_{\rm{f}} \simeq 0.1 c$.  
\end{itemize}

Other two models have been proposed in the literature with features similar to the ones of the scenarios described above for what concerns the neutrino production. 
\cite{Lyutikov:2022xri} suggests that LFBOTs arise from the accretion induced collapse  of a binary white dwarf merger. In this case, neutrinos may be produced at the highly magnetized and highly relativistic wind termination shock, responsible for the observed radio emission. In this scenario, we expect a neutrino signal  similar to the one of the cocoon model (from CSM interaction only), because of the similarity with the model parameters considered in~\cite{Lyutikov:2022xri}.
\cite{Soker:2022gna} invokes a common envelope  phase between a red supergiant  and a CO. 
This mechanism shares common features with the one proposed in~\cite{Gottlieb:2022old}. Nevertheless, while the former predicts baryon loaded jets, the latter  invokes relativistic jets. 
Neutrino production from the jet model proposed in~\cite{Soker:2022gna} may mimic the results obtained in~\cite{Grichener:2021xeg}.
Moreover, as for the scenario of~\cite{Metzger:2022xep},  a common envelope phase,  during which an asymmetric CSM  forms, is proposed. The parameters obtained in the common envelope jet SN impostor scenario are similar to the cocoon model as for the total energy and mass of the ejecta, as well as for the CSM properties.  
Results similar to the ones of the cocoon model should hold for the common envelope jet SN impostor scenario, when taking into account CSM interaction. 
Hence, in the following we focus on the cocoon and merger models only.

Independently of its origin, the fast outflow  propagates  outwards  in the surrounding CSM, giving rise to the observed radio spectrum.
Observations suggest a certain degree of asymmetry in the LFBOTs outflows~\citep{Margutti:2018rri, Coppejans:2020nxp, Yao:2021qqz}. Nevertheless, for the sake of simplicity, we consider a spherically symmetric geometry both for the ejecta and the CSM. 
We parametrize the CSM with a wind profile
\begin{equation}
n_{p, \rm{CSM}} (R) = \frac{\dot{M}}{4 \pi m_p v_w R^2} \; ,
\label{eq:densityCSM}
\end{equation}
where ${\dot{M}}$ is the mass-loss rate of the star and $v_w$ is the wind velocity. The CSM extends up to $R_{\rm{CSM}}$ and its mass is obtained by integrating Eq.~\ref{eq:densityCSM} over the volume of the CSM shell, ${\rm{d}} V_{\rm{CSM}} = 4 \pi R^2 {\rm{d}} R$. Note however that radio observations of AT2018cow indicate a steeper density profile for the CSM, see e.g.~\cite{AJ:2021kic}. Here, we assume a standard wind profile for a general case.

As the outflow expands in the CSM,  forward and  reverse shocks form---propagating in the stellar wind and back to the ejecta in mass coordinates, respectively. 
Both the forward and  reverse shocks contribute to  neutrino production. On the basis of similarities with the SN  scenario, the forward shock is expected to be the main dissipation site of the kinetic energy of the outflow~\citep[e.g.,][]{Ellison:2007bga, Patnaude:2008gq,2010MNRAS.406.2633S,Suzuki:2020qui, 2014ApJ...783...33S,2018ApJ...853...46S}; hence,  we focus on  the forward shock only, which moves with  speed $v_{\rm{sh}} \simeq v_{\rm{f}}$. 

If the outflow expands in a dense CSM with optical depth $\tau_{\rm{CSM}}$, the forward shock is radiation mediated as long as $\tau_{\rm{CSM}} \gg 1$ and particle acceleration is not efficient~\citep{Levinson:2007rj, Katz:2011zx, Murase:2010cu}. Radiation escapes at the breakout radius $R_{\rm{bo}}$,  when the optical depth drops below $v_{\rm{sh}}/c$. The breakout radius is  obtained by solving the following equation:
\begin{equation}
\tau_{\rm{CSM}}= \int_{R_{\rm{bo}}}^{R_{\rm{CSM}}} dr\ \sigma_T n_{p, \rm{CSM}}(R) = \frac{c}{v_{sh}} \; .
\label{eq:breakoutRad}
\end{equation}
Existing data suggest that  the LFBOT ejecta were possibly slowly decelerating during the time of observations~\citep[e.g.,][]{Coppejans:2020nxp}. Nevertheless, this behavior is not well probed and the treatment of  deceleration of a mildly-relativistic blastwave is not straightforward~\citep{Coughlin:2019liv}. Hereafter, we assume that the shock freely moves with  constant speed $v_{\rm{sh}}$ up to the deceleration radius
\begin{equation}
    R_{\rm{dec}}= R_{\rm{bo}}+\frac{M_{\rm{ej}}}{4 \pi m_p n_{p, \rm{bo}} R_{\rm{bo}}^2 } \; ,
 \label{eq:decRadius}
\end{equation}
where $M_{\rm{ej}}$ is the mass of the ejecta and $n_{p, \rm{bo}} = n_{p, \rm{CSM}}(R_{\rm{bo}})$. At this radius, the ejecta have swept-up a mass comparable to $M_{\rm{ej}}$ from the CSM.

\subsubsection{Proton energy distribution}
Diffusive shock acceleration of the CSM protons occurs at $ R \gtrsim R_{\rm{bo}}$ and  accelerated protons are assumed to have a power-law energy distribution. For a wind-like CSM, the proton distribution reads [in units of GeV$^{-1}$ cm$^{-3}$]
\begin{equation}
	\tilde{n}_{{p}}(\tilde{E}_{p}) \equiv \frac{{\rm{d}}^{2} \tilde{N}_{{p}}}{{\rm{d}}\tilde{E}_{{p}}{\rm{d}} \tilde{V}}  = \tilde{A}_p \tilde{E}_{{p}}^{-k_p} \Theta(\tilde{E}_{{p}} - \tilde{E}_{p, \rm{min}}) \Theta(\tilde{E}_{p, \rm{max}} - \tilde{E}_{\rm{p}})\ ;
\label{eq:protonInjection}
\end{equation}
as for the choked jet scenario, we fix the proton spectral index $k_p = 2$. Moreover, the minimum energy of protons is $\tilde{E}_{p, \rm{min}} = m_p c^2$, since these shocks  are not relativistic. The maximum energy of shock-accelerated protons is fixed by the condition that the acceleration timescale is shorter than the total cooling timescale, i.e. $\tilde{t}^{-1}_{\rm{acc}} \le \tilde{t}^{-1}_{\rm{cool}}$ (see Appendix~\ref{sec:A}). Note that for CSM interaction there is no difference between the comoving frame of the shock and the CO frame, since the involved speeds  are sub-relativistic. Hence, the primed quantities are equivalent to the twiddled ones. 

$\tilde{A}_p = {9 \epsilon_{{p}} n_{{p, \rm{CSM}}}}(R) m_p c^2/[{8  {\rm{ln}} (\tilde{E}_{p, \rm{max}}/ \tilde{E}_{p, \rm{min}})}] ({v_{\rm{sh}}}/c)^2 $ is the normalization constant. Here, $\epsilon_p$ is the fraction of the post-shock internal energy, $\tilde{e}_{\rm{th}} = 9 m_{{p}} c^2 (v_{\rm{sh}}/c)^2 n_{p, \rm{CSM}} (R)/ 8$, that goes in accelerated protons. The fraction $\epsilon_B$ of $\tilde{e}_{\rm{th}}$ is instead stored in the magnetic field generated at the forward shock: $\tilde{B} = \sqrt{9 \pi \epsilon_B m_p c^2 (v_{\rm{sh}}/c)^2 n_{p, \rm{CSM}}(R)}$. We stress that the quantities introduced so far for CSM interaction evolve with the radius of the expanding outflow, and hence with time. 

Electrons are expected to be accelerated together with protons at the forward shock and produce the synchrotron self-absorption spectrum observed in the radio band. The electron population responsible for the radio emission is still under debate~\citep{Ho:2021fyb,Margalit:2021kuf}. Nevertheless, we verified that $p \gamma$ interactions are negligible for a wide range of parameters, consistently with the results reported in~\cite{Murase:2010cu,Fang:2020bkm}. Hence, we do not introduce any photon distribution and neglect neutrino production through  $p \gamma$ interactions in the context of CSM-ejecta interaction (see Sec.~\ref{sec:neutrinoProduction}).

\section{Benchmark luminous fast blue optical transients: AT2018cow and CSS161010}\label{sec:parameters}
\begin{table*}[!ht]
\caption{Benchmark input parameters characteristic of AT2018cow and CSS161010 adopted in this work. Some parameters are inferred from  observations, while others denote  typical values derived on theoretical grounds or combining observations and theoretical arguments. The following references are quoted in the table: [1]~\cite{Prentice:2018qxn}, [2]~\cite{Coppejans:2020nxp}, [3]~\cite{Perley:2018oky}, [4]~\cite{Granot:2006hf}, [5]~\cite{Kumar:2014upa}, [6]~\cite{AJ:2021kic}, [7]~\cite{Margutti:2018rri}, [8]~\cite{Gottlieb:2022old}, [9]~\cite{1969ApJ...157.1395O}, [10]~\cite{Mizuta:2013yma}, [11]~\cite{Meszaros:2001ms}, [12]~\cite{Tan:2000nz}, [13]~\cite{Kobayashi:1997jk}, [14]~\cite{Guetta:2000ye}, [15]~\cite{2011ApJ...726...75S}, [16]~\cite{He:2018lwb}, [17]~\cite{kippenhahn1990stellar}, [18]~\cite{Ho:2018emo}, [19]~\cite{Caprioli:2013dca}, [20]~\cite{Metzger:2022xep}. }
		\begin{flushleft}
		\begin{tabularx}{\textwidth}{ccccc}
		\toprule
		\toprule
		{Parameter} & {Symbol} & {AT2018cow} & {CSS161010} & {References}\\
		\toprule
		Luminosity distance & $d_L$ & $60$~Mpc & $150$~Mpc  & [1, 2] \\
		Declination & $\delta$ & $22^\circ$ & $-8^\circ$ & [2, 3] \\
		\toprule
		  & choked jet  & & &  \\
		\toprule
		Opening angle & $\theta_j$ & 0.2 & 0.2 & [4, 5, 6] \\
		Isotropic energy & $\tilde{E}_{j}^{\rm{iso}} \; (\rm{erg})$ & $10^{50}$--$10^{52}$  & $10^{50}$--$10^{52}$ & [2, 7, 8] \\
		Jet lifetime & $ \tilde{t}_j$~(s) & $10$--$10^6$ & $10$--$10^6$  & [9, 17] \\ 
		Lorentz factor & $\Gamma $ & $10$--$100$ & $10$--$100 $  & [10, 11, 12] \\
		Dissipation efficiency (IS) & $\epsilon_{\rm{d}} $ & $0.2$ & $0.2$  & [13, 14] \\
		Accelerated proton energy fraction (IS) & $\epsilon_{\rm{p}} $ & $0.1$ & $0.1$  & [15] \\
		Magnetic energy density fraction (IS) & $\epsilon_{\rm{B}}$ & $0.1$ & $0.1$  & [15] \\
		Accelerated electron energy fraction (RS) & $\epsilon_e^{\rm{RS}}$ & $0.1$ & $0.1$ & [16] \\
		\toprule
		 & CSM interaction, cocoon model   & & &\\
		\toprule
		Fast outflow velocity & $v_{\rm{f}} $	& $0.2 c$ & $0.55 c$ & [2, 6, 7, 18] \\
		Ejecta energy & $\tilde{E}_{\rm{ej}}\; (\rm{erg})$ & $ 4 \times 10^{48}$--$10^{51}$  & $6 \times 10^{49}$--$10^{51}$  & [2, 7, 18] \\
		Mass-loss rate & $\dot{M} \; (M_\odot \; \rm{yr}^{-1})$ & $ 10^{-4}$--$10^{-3}$ & $10^{-4}$--$10^{-3}$  & [2, 7, 18] \\
		Ejecta mass & $M_{\rm{ej}} (M_\odot) $ & $ 1 \times 10^{-4}$--$3 \times 10^{-2}$	& $ 2.2 \times 10^{-4}$--$4 \times 10^{-3}$ & [2, 7, 18]\\
		Wind velocity & $v_{w}$~(km s$^{-1}$) & 1000 & 1000 & [2, 7, 18] \\
		CSM radius & $R_{\rm{CSM}}$~(cm) & $1.7 \times 10^{16}$ & $3 \times 10^{17}$ & [2, 18]\\
		Accelerated proton energy fraction & $\epsilon_p$ & $0.1$ & $0.1$ & [19]\\
		Magnetic energy density fraction & $\epsilon_B$ & $0.01$ & $0.01$ & [2, 6, 7,  18] \\
		\toprule
		 & CSM interaction, merger model  & & & \\
		\toprule
		Fast outflow velocity & $v_{\rm{f}} $	& $0.2 c$ & 0.55 c & [2, 6, 7, 18] \\
		Ejecta energy & $\tilde{E}_{\rm{ej}} \; (\rm{erg})$ & $4 \times 10^{48}$--$10^{51}$  & $6 \times 10^{49}$--$10^{51}$ & [2, 7, 20] \\
		Mass-loss rate & $\dot{M} \; (M_\odot \; \rm{yr}^{-1})$ & $7\times 10^{-6}$--$7 \times 10^{-5}$ & $7 \times 10^{-6}$--$7 \times 10^{-5}$ & [2, 7, 20] \\
		Ejecta mass & $M_{\rm{ej}} (M_\odot) $ & $ 10^{-4}$--$3 \times 10^{-2}$ &  $ 2.2 \times 10^{-4}$--$4 \times 10^{-3}$  & [2, 7, 20]  \\
		Wind velocity & $v_{w}$~(km s$^{-1}$) & 10 & 10 & [20] \\
		CSM radius & $R_{\rm{CSM}}$~(cm) & $3 \times 10^{16}$   & $3 \times 10^{16}$ & [20]\\
		Accelerated proton energy fraction & $\epsilon_p$ & $0.1$ & $0.1$ & [19]\\
		Magnetic energy density fraction & $\epsilon_B$ & $0.01$ & $0.01$  & [2, 6, 7, 18] \\		
		\toprule
		\toprule
	\end{tabularx}
	\end{flushleft}
\label{table:parametersA} 	
\end{table*}

\begin{figure*}[t]
\hspace{-0.8cm}
	\includegraphics[width=0.55\textwidth]{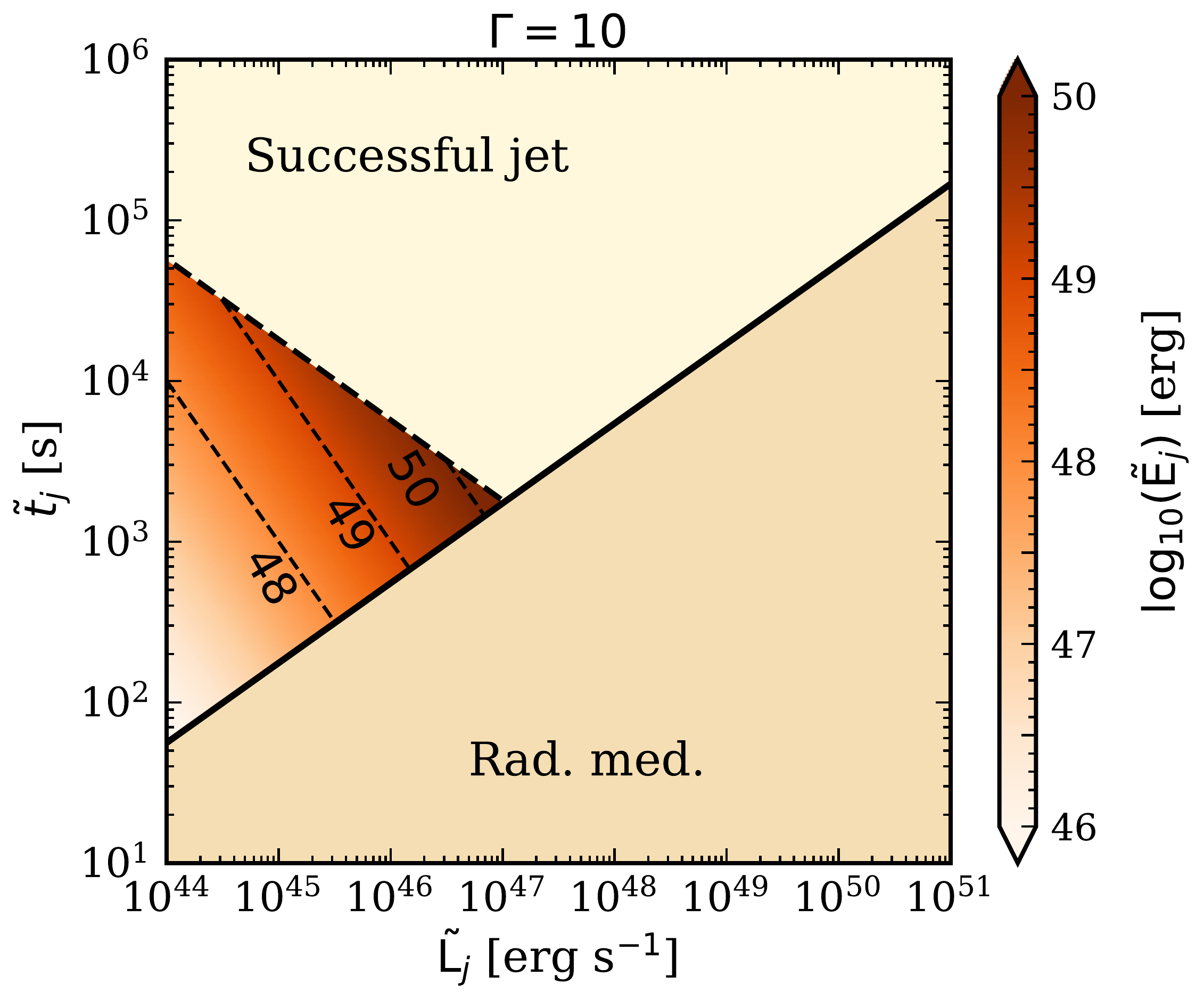}
	\includegraphics[width=0.55\textwidth]{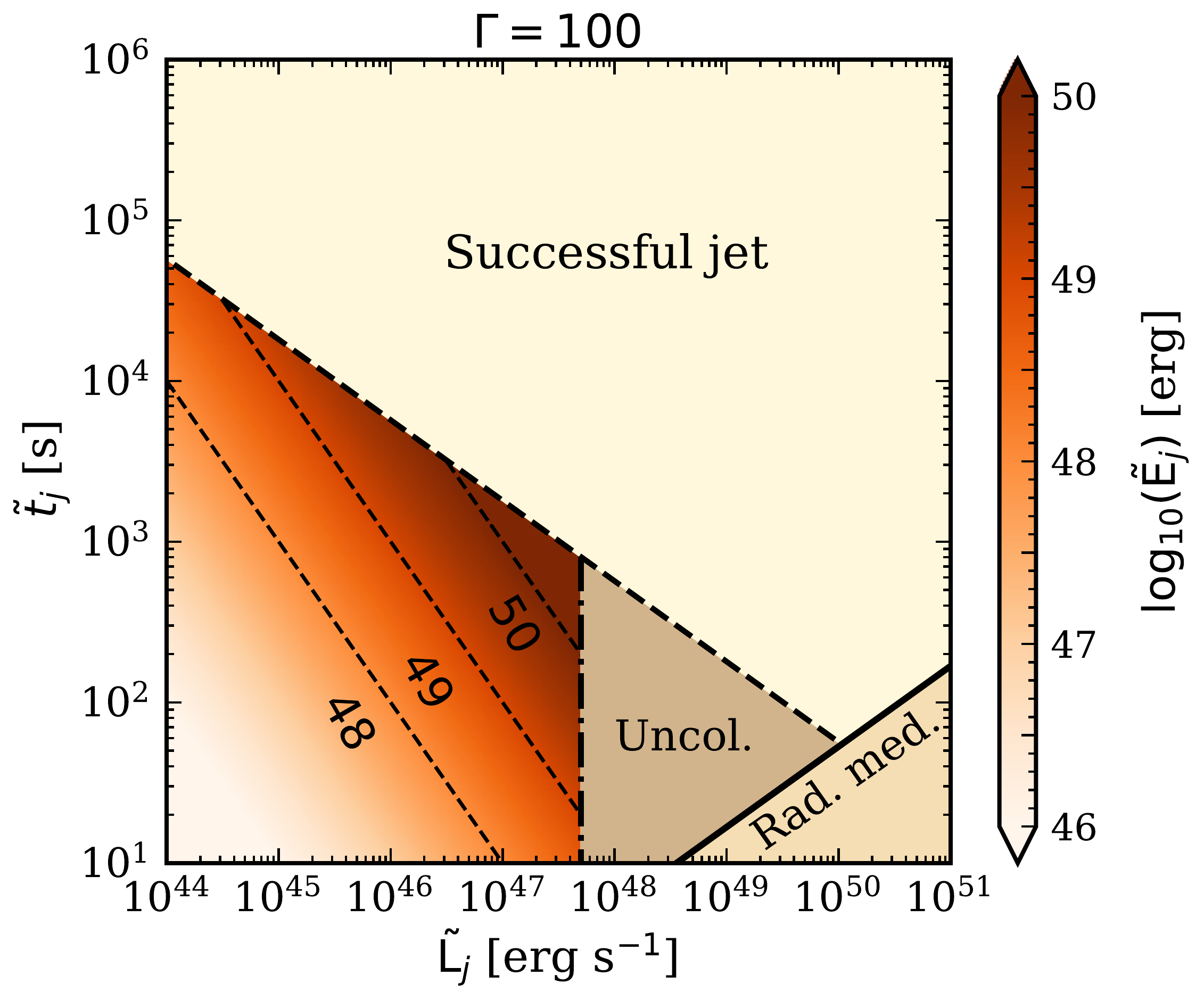}
	\caption{Contour plot of the  energy  injected in the jet by the central engine ($\tilde{E}_{j} = \tilde{L}_j \tilde{t}_j$) in the plane spanned by $\tilde{L}_{j}$ and $\tilde{t}_j$ for $\Gamma = 10$  (left panel) and $\Gamma = 100$ (right panel).  The light yellow region is excluded since it would give rise to a successful jet. The light-brown region in the right lower corner is excluded because the jet would be radiation mediated (``Rad.~med.''; see Eq.~\ref{eq:rad_condition}). For $\Gamma = 100$, we exclude an additional region corresponding to an uncollimated jet (``Uncol''; brown region in the right panel). In the allowed region of the parameter space, the black-dashed lines are meant to guide the eye and correspond to $\tilde{E}_j = 10^{48}, 10^{49}, 10^{50}$~erg.}
	\label{fig:parameterSpace}
\end{figure*}
In this section, we provide an overview on the  parameters characteristic of AT2018cow and CSS161010.
We select these two transients as representative of the  detected LFBOTs for two reasons. First, they are the closest ones ($d_L \simeq 60$~Mpc for AT2018cow and $d_L \simeq 150$~Mpc for CSS161010; $d_L$ is the luminosity distance,  defined as in Sec.~\ref{sec:fluxEarth}); second, while these two LFBOTs share similar  CSM densities, extension of the CSM, ejecta mass and kinetic energy  as the population of LFBOTs, their fastest ejecta span the entire range of values inferred. AT2018cow showed $v_{\rm{f}} \simeq 0.1$--$0.2 c$~\citep{Margutti:2018rri,Ho:2018emo, AJ:2021kic}, while CSS161010 is the fastest LFBOT observed to date with $v_{\rm{f}} \simeq 0.55 c$ \citep{Coppejans:2020nxp}.
We fix the speed of the fastest component of the outflow as measured from observations. The other characteristic parameters are still uncertain,  hence we vary them within an uncertainty range. The parameters adopted for the choked jet (opening angle $\theta_j$, Lorentz factor $\Gamma$, and lifetime $\tilde{t}_j$) are fixed on the basis of  theoretical arguments as justified below. The typical parameters adopted for the choked jet and for CSM interaction are summarized in Table~\ref{table:parametersA}. 

As for  the cocoon model harboring a choked jet, $\tilde{E}_{ j} = \tilde{L}_{j} \tilde{t}_j$ corresponds to the physical energy injected by the central engine into the jet, whose opening angle is assumed to be $\theta_j = 0.2$~rad (e.g.~\citealt{Granot:2006hf, Kumar:2014upa}). Since the jet is choked, all of its energy is transferred to the cocoon, i.e.~the cocoon breaks out with energy $\tilde{E}_{\rm{ej}} \simeq \tilde{E}_{j}$; note that, in principle, we should consider that a fraction of the jet energy is dissipated at the internal shocks, nevertheless this fraction is small enough  to be negligible [$\sim 10 \%$~\citep{Kobayashi:1997jk}]. The kinetic energy $\tilde{E}_{k}$ of the ejecta interacting with the CSM has been estimated from the radio data and it represents a lower limit on the total energy of the outflow, $\tilde{E}_{\rm{ej}}$ (see ``CSM interaction, cocoon model'' in Table~\ref{table:parametersA}). The upper limit on the total energy of the outflow is not directly inferred from observations, but  estimations of its range of variability have been attempted.  Thus, we vary the energy injected in the jet in the interval spanned by  the lower and upper limits of the outflow energy, obtained by combining observations and theoretical assumptions  (see ``choked jet'' in Table~\ref{table:parametersA} and references therein). As mentioned in Sec.~\ref{sec:chokedjet}, the dynamics of the jet is conveniently described by the isotropic equivalent quantities;  we refer to the isotropic equivalent energy of the jet: $\tilde{E}_j^{\rm{iso}}= \tilde{E}_j/(\theta_j^2/4)$.

The Lorentz factor of the jet is not measured. Hence, we rely on  two extreme cases: $\Gamma = 10$ and $100$. This choice  is due to the fact that  numerical simulations and semi-analytical models suggest that the jet propagates in the stellar core with $\Gamma \simeq 1$--$10$~\citep{Mizuta:2013yma, Harrison:2017jvs}. Nevertheless, when the jet pierces the stellar core at $R_\star \simeq 10^{11}$~cm and enters the extended envelope, it may be accelerated up to $\Gamma \lesssim 100$ because of the sudden drop in  density~\citep{Meszaros:2001vr, Tan:2000nz}. 

The jet lifetime is linked to the CO physics. The CO harboring relativistic jets can  be either a black hole~\citep{Gottlieb:2021srg, Quataert:2018gnt} or a millisecond magnetar~\citep{2011MNRAS.413.2031M}.
If we assume that the central engine of LFBOTs is a magnetar with initial spin period $P_i$, magnetic field $B_m$ and mass $M_m = 1.4 M_\odot$ then the upper limit on the jet lifetime is set by the spin-down period~\citep{1969ApJ...157.1395O}:
\begin{equation}
\tilde{t}_{\rm{sd}} = 2.0 \times 10^5 \; \rm{s} \; \biggl( \frac{P_i}{10^{-3} \; \rm{s}} \biggr)^2 \biggl( \frac{B_m}{10^{14} \; \rm{G}} \biggr)^2 \; .
\end{equation} 
Following~\cite{Fang:2018hjp}, for $P_i = 10$~ms and $B_m = 10^{15}$~G,  we obtain $\tilde{t}_j \lesssim \tilde{t}_{\rm{sd}} = 2 \times 10^5$~s.
If  the CO  is a black hole,  the upper limit on the jet lifetime is set by the free-fall time of the stellar material~\citep{kippenhahn1990stellar}:
\begin{equation}
    \tilde{t}_{\rm{ff}} \simeq 1.7 \times 10^7 \; \rm{s} \; \left( \frac{R_{\rm{BH}}}{10^{13.5} \; \rm{cm}} \right)^{3/2} \left(\frac{M_{\rm{BH}}}{M_\odot}\right)^{-1/2}\ ,
\end{equation}
where $M_{\rm{BH}}$ is the black hole mass and $R_{\rm{BH}}$ the distance from it. Since the nature of the CO powering LFBOTs as well as the presence of a jetted outflow are  uncertain,  we vary the jet lifetime in $\tilde{t}_j \in \bigl[10 , 10^6  \bigr]$~s. Note, however, that a short lifetime ($\tilde{t}_j < 10^3$~s) may require an amount of energy released by the CO larger than the sum of the observed radiated energy and the kinetic energy of the ejecta. This consideration arises when extrapolating the X-ray light-curve---likely associated with the CO powering LFBOTs~\citep[e.g.][]{Margutti:2018rri, Coppejans:2020nxp}---back to early times ($\tilde{t} \sim \tilde{t}_j$). Nevertheless, there is no robust signature that allows to confidently exclude  shorter CO lifetimes. Hence, we choose to span a wide range for $\tilde{t}_j$. Finally, the microphysical parameters $\epsilon_{B}$, $\epsilon_{p}$, and $\epsilon_{e}^{\rm{RS}}$ are fixed to typical values of choked jets; see ``choked jet'' in Table~\ref{table:parametersA} and references therein. 

Note that the same energy $\tilde{E}_j$ can be injected from the CO for different $(\tilde{L}_{j}, \tilde{t}_j)$ pairs. Since our main goal is to explore viable mechanisms for neutrino production in LFBOTs,   not all  $(\tilde{L}_{j}, \tilde{t}_j)$ pairs are allowed, as shown in Fig.~\ref{fig:parameterSpace}. In fact, the  $(\tilde{L}_{j}, \tilde{t}_j)$ pairs that do not satisfy, simultaneously,  the choked jet condition ($R_h < R_{\rm{ext}}$, with $R_h$ given by Eq.~\ref{eq:head_radius}) as well as the acceleration constraint in Eq.~\ref{eq:rad_condition} are excluded. Examples of the allowed $(\tilde{L}_{j}, \tilde{t}_j)$ pairs are shown in Fig.~\ref{fig:parameterSpace} for $\Gamma =10$ and $100$. We also exclude the $(\tilde{L}_{j}, \tilde{t}_j)$ pairs  leading to an uncollimated jet in the extended envelope for the fixed $\theta_j$, as suggested by numerical simulations and  implied by observations~\citep{Gottlieb:2022old},  see~\cite{2011ApJ...740..100B, Xiao:2014vga} for details\footnote{We assume a density profile of the stellar core $\rho_{\rm{star}} (R) = M_\star /(4 \pi R_\star) R^{-2}$, valid up to the He envelope; this profile follows~\cite{Matzner:1998mg, Xiao:2014vga} for progenitors harboring choked jets. For the mass of the stellar core and its radius we use $M_\star = 4 M_\sun$ and $R_\star = 6 \times 10^{11}$~cm, respectively, inspired by~\cite{Gottlieb:2022old} that reproduces the lightcurve of AT2018cow.}. Uncollimated outflows are ruled out by energetic considerations, since they would require a total energy of the ejecta, $\tilde{E}_{\rm{ej}} \simeq 10^{53}$~erg, much larger than the one estimated for LFBOTs, i.e.~$\tilde{E}_{\rm{ej}}  \simeq 10^{50}$--$10^{51}$~erg~\citep{Coppejans:2020nxp, Ho:2020hwf, Ho:2018emo}.
In Fig.~\ref{fig:parameterSpace}, we  consider  isocontours of the isotropic energy $\tilde{E}_{j}$ in the $(\tilde{L}_{j}, \tilde{t}_j)$ parameter space .  Note that, for $\Gamma = 10$, the region excluded from the collimation argument overlaps with the area already excluded; therefore, we do not show it explicitly. 

Concerning CSM interaction occurring in the cocoon model, if $v_{\rm{f}}$ is the speed of the fastest component of the cocoon responsible for the observed radio emission and $E_k = \tilde{E}_{k}/(1+z)$ its kinetic energy,  its mass $M_{\rm{ej}}$ can be obtained through  the following relation
 \begin{equation}
 v_{\rm{ej}} = \sqrt{\frac{2 E_{k}}{M_{\rm{ej}}}} \ .
 \label{eq:vej}
 \end{equation}
We then vary $M_{\rm{ej}}$ in the range  corresponding to the upper and lower limits on the kinetic energy of the outflow. The former is obtained by assuming that all the energy of the ejecta is converted into kinetic energy, i.e.~$\tilde{E}_{k} = \tilde{E}_{\rm{ej}}$; the latter is constrained from observations. The range of variability of $M_{\rm{ej}}$  is shown in Table~\ref{table:parametersA} for AT2018cow and CSS161010 (see under ``CSM interaction, cocoon model''). 
The mass-loss rate $\dot{M}$ spans the range hinted from radio data, while the CSM radius  is fixed from the latest radio observations; see ``CSM interaction, cocoon model'' in Table~\ref{table:parametersA} and references therein.  

For the merger model, we fix the upper limit on the total energy of the ejecta at the theoretical value estimated by \cite{Metzger:2022xep}. We instead vary  the mass of the fast ejecta by using Eq.~\ref{eq:vej}, following the argument reported above concerning the upper and lower limits on the kinetic energy. Finally, the mass-loss rate spans a range obtained from theoretical predictions of the model, while the extension of the CSM is fixed from theoretical estimations~\citep{Metzger:2022xep}. All the aforementioned parameters and their variability ranges are listed in the section ``CSM interaction, merger model'' of Table~\ref{table:parametersA}.

\section{Neutrino production}\label{sec:neutrinoProduction}
In this section, we summarize the viable mechanisms for  neutrino production in LFBOTs. In particular, we discuss interactions between shock accelerated protons and target photons at the internal shocks ($p \gamma$ interactions) in the choked jet  and interactions between shock-accelerated protons and a steady target of protons ($pp$ interactions), taking place when the outflow expands in the  CSM. In both cases, we  present the procedure adopted to compute the high-energy neutrino flux at Earth. 

\subsection{Neutrino production via  proton-photon interactions}
Protons accelerated at the internal shocks  interact with thermal photons escaping from the jet head and going back to the unshocked jet. Efficient $p \gamma$ interactions take place at the internal shock, mainly through the $\Delta^+$ channel
\begin{equation}
   p+ \gamma \longrightarrow \Delta^+ \longrightarrow
   \begin{system}
   n + \pi^+ \; \; \; \; \; \; \; 1/3 \; \rm{\; of\; all\; cases} \\
   p+   \pi^0  \; \; \; \; \; \; \;   2/3 \; \rm{\; of \; all\; cases }  \ , 
   \end{system}
    \label{reaction_channel}
\end{equation} 
while we can safely neglect $pp$ interactions at the internal shocks, since they are subleading (see Appendix~\ref{sec:A}). The reaction channel in Eq.~\ref{reaction_channel} is followed by the decay of neutral pions into photons: $\pi^0 \longrightarrow  2 \gamma$. At the same time, neutrinos can be copiously produced in the decay chain $\pi^+ \longrightarrow \mu^ + + \nu_\mu$, followed by the muon decay $\mu^+ \longrightarrow \bar{\nu}_\mu + \nu_e + e^+$. 

We rely on the photo-hadronic model presented in~\cite{Hummer:2010vx}.  Hence, given the injected energy distribution of protons $[n^\prime_p(E^\prime_p)]$ and the distribution of target photons  $[n^\prime_\gamma(E^\prime_\gamma)]$, the rate of production of secondary particles $l$ (with $l= \pi^{\pm}, \pi^0, K^+$) in the comoving frame of the unshocked jet is given by [in units of GeV$^{-1}$~cm$^{-3}$~s$^{-1}$]:
\begin{equation}
Q^\prime_l(E^\prime_l) = c \int_{E^\prime_l}^\infty \frac{d E^\prime_p}{E^\prime_p} n^\prime(E^\prime_p) \int_{E_{\rm th}/2 \gamma^\prime_p}^\infty dE^\prime_\gamma n^\prime_\gamma(E^\prime_\gamma) R(x, y)\ ,
\label{eq:rate_secondaries}
\end{equation}
where $x = E^\prime_l/E^\prime_p$ is the fraction of proton energy which is given to secondary particles, $y = \gamma^\prime_p E^\prime_l$ and $R(x, y)$ is the response function, which contains the physics of the interaction. 
The initial distributions of protons and photons are given by Eqs.~\ref{eq:protonJET} and \ref{eq:BBspectrum}, respectively.

Before decaying, each charged meson $l$ undergoes energy losses, parametrized through the cooling time $t^{\prime -1}_{l, \rm{cool}}$, see Appendix~\ref{sec:A}. Therefore, the spectrum at the decay is
\begin{equation}
Q^{\prime \rm{dec}}_l(E^\prime_l) = Q^\prime_{l}(E^\prime_{l}) \biggr[1 - \exp\biggl(- \frac{t^\prime_{l, \rm{cool}} m_l}{E^\prime_l \tau^\prime_l}\biggr)\biggr]\  ,
\label{eq:decayed_spectrum}
\end{equation}
where $\tau^\prime_l$ is the lifetime of the meson $l$. The comoving neutrino spectrum from decayed mesons is [in units of GeV$^{-1}$~cm$^{-3}$~s$^{-1}$]:
\begin{equation}
Q^\prime_{\nu_\alpha}(E^\prime_{\nu}) = \int_{E^\prime_{\nu}}^{\infty} \frac{dE^\prime_l}{E^\prime_l} Q^{^\prime \rm{dec}}_{l}(E^\prime_l) F_{l \rightarrow \nu_\alpha} \biggl(\frac{E^\prime_{\nu}}{E^\prime_l} \biggr)\ ,
\label{eq:rate_neutrini}
\end{equation}
where $\alpha = e, \mu$ is the neutrino flavor at production and $F_{l \rightarrow \nu_\alpha}$ is provided in~\cite{Lipari:2007su}. We use $\nu_\alpha \equiv \nu_\alpha + \bar{\nu}_\alpha$, i.e.~we do not distinguish between neutrinos and antineutrinos.

 Magnetic fields in the internal shock are not large enough to efficiently cool kaons, that have a larger mass and a shorter lifetime compared to pions and muons. Therefore, they suffer less energy losses and do not contribute significantly to the neutrino spectrum, even though they may become important at high energies~\citep{He:2012tq, Asano:2006zzb, Petropoulou:2014lja, Tamborra:2015qza}. 

\vspace{0.5cm}
\subsection{Neutrino production via proton-proton interactions}
Similar to  SNe, stellar outflows interacting with dense CSM  can be neutrino factories~\citep{Murase:2010cu, Pitik:2021dyf, Petropoulou:2017ymv, Petropoulou:2016zar, 2011arXiv1106.1898K, Murase:2013kda, Cardillo:2015zda,Zirakashvili:2015mua, Murase:2020lnu,Sarmah:2022vra}, when  protons accelerated at the forward shock between the ejecta and the CSM interact with the steady target protons of the CSM.

Given the  population of injected shock-accelerated protons in Eq.~\ref{eq:protonInjection}, the proton distribution evolves as~\citep{1997ApJ...490..619S, 2012ApJ...751...65F, Petropoulou:2016zar}:
\begin{equation}
	\frac{\partial \tilde{N}_{\rm{p}}(\tilde{\gamma}_{\rm{p}}, R)}{\partial R} - \dfrac{\partial }{\partial \tilde{\gamma}_{\rm{p}}}\bigg[\frac{\tilde{\gamma}_{\rm{p}}}{R} \tilde{N}_{\rm{p}}(\tilde{\gamma}_{\rm{p}}, R)\bigg] + \frac{\tilde{N}_{\rm{p}}(\tilde{\gamma}_{\rm{p}}, R)}{v_{\rm{sh}}\tilde{t}_{pp}(R)} = \tilde{Q}(\tilde{\gamma}_p)\ ,
\label{eq:Np}	
\end{equation}
where $\tilde{N}_{\rm{p}}(\tilde{\gamma}_{\rm{p}},R)$ is the total number of protons with Lorentz factor between $\tilde{\gamma}_p$ and $\tilde{\gamma}_p + d \tilde{\gamma}_p$ contained in the shell of shocked material at radius $R $ and $\tilde{Q}(\tilde{E}_p) = \pi R_{\rm{bo}}^2 \tilde{n}(\tilde{E}_p/m_p c^2, 
R=R_{\rm{bo}})/ (m_p c^2)$ is the proton injection rate at the breakout radius [in units of cm$^{-1}$]. The second term on the left-hand side of Eq.~\ref{eq:Np} parametrizes  adiabatic losses due to the expansion of the shocked shell, while the third term corresponds to $pp$ collisions, treated as an escape term~\citep{1997ApJ...490..619S}.

The neutrino production rates for neutrinos of flavor $\alpha$, $ Q_{\nu_{\alpha}}$ are given by  [in units of {GeV}$^{-1}$ {cm}$^{-1}$]~\citep{Kelner:2006tc}:
\begin{eqnarray}
\label{eq:Q_nu}
	\tilde{Q}_{\nu_{\alpha}}(\tilde{E}_{\nu}, R) &=& \frac{4 n_{p, \rm{CSM}}(R) m_{{p}} c^{3}}{v_{\rm{sh}}} \int_{0}^{1} dx \frac{\sigma_{{pp}}(\tilde{E}_{\nu}/x)}{x}  \\  \nonumber &  & \times \;  \tilde{N}_{\rm{p}} \bigg(\frac{\tilde{E}_{\nu}}{x m_{\rm{p}} c^{2}}, R\bigg) F_{\nu_\alpha} (\tilde{E}_\nu, x)\ ,
\end{eqnarray}
where $ x = \tilde{E}_{\nu}/\tilde{E}_{\rm{p}} $ and  the function $F_{\nu_{\alpha}} $ is provided in~\cite{Kelner:2006tc}. Note that Eq.~\ref{eq:Q_nu} is only valid  for $ E_{\rm{p}} > 0.1$~TeV, which is the energy range we are interested in.

\subsection{Neutrino flux at Earth}\label{sec:fluxEarth}
On their way to Earth, neutrinos undergo flavor conversion. The  observed distribution for the flavor $\nu_\alpha$ (with $\alpha=e, \mu, \tau$) is [GeV$^{-1}$~cm$^{-2}$~s$^{-1}$]
\begin{equation}
F_{\nu_\alpha}(E_{\nu}, z) = \mathcal{T} \frac{(1+z)^2}{4 \pi d_L^2(z)} \sum_\beta P_{\nu_\beta \rightarrow \nu_\alpha} (E_{\nu}) \mathcal{Q}^{\prime}_{\nu_\beta}\left({E_{\nu} \mathcal{L}}\right)\ ,
\label{eq:neutrino_flux}
\end{equation}
with $\mathcal{Q}^{\prime}_{\nu_\beta}\left({E_{\nu} \mathcal{L}}\right)$  being the neutrino production rate in the comoving jet ($p \gamma$ interactions) or in the center of explosion ($pp$ interactions) frame, given by Eqs.~\ref{eq:rate_neutrini} and \ref{eq:Q_nu}, respectively. The  constant  $\mathcal{T} = V^\prime_{\rm{iso}} = 4 \pi R_{\rm{IS}}^3 /(2 \Gamma)$ represents the isotropic volume of the interaction region~\citep{Baerwald:2011ee} in the choked jet scenario, while $\mathcal{T} = v_{\rm{sh}}$ for CSM-ejecta interaction. Note that $\mathcal{T}$ has different dimensions in the choked jet scenario compared to the CSM-ejecta interaction case, because of the different dimensionality of the corresponding neutrino injection rates, see Eqs.~\ref{eq:rate_neutrini} and~\ref{eq:Q_nu}. Moreover, the Lorentz conversion factor is $\mathcal{L} = (1+z)/ \Gamma$ for the choked jet and  $\mathcal{L} = (1+z)$ for CSM interaction. The neutrino oscillation probability $P_{\nu_\beta \rightarrow \nu_\alpha}= P_{\bar{\nu}_\beta \rightarrow \bar{\nu}_\alpha}$ is given by~\citep{Anchordoqui:2013dnh,Farzan:2008eg}: 
\begin{eqnarray}
P_{\nu_e \rightarrow \nu_\mu} &=& P_{\nu_\mu \rightarrow \nu_e} = P_{\nu_e \rightarrow \nu_\tau} = \frac{1}{4} \sin^2 2\theta_{12}\ , \\
P_{\nu_\mu \rightarrow \nu_\mu} &=& P_{\nu_\mu \rightarrow \nu_\tau  }= \frac{1}{8}(4-\sin^2 \theta_{12})\ ,\\
P_{\nu_e \rightarrow \nu_e} &=& 1- \frac{1}{2} \sin^2 2\theta_{12}\ ,
\end{eqnarray}
with $\theta_{12} \simeq 33.5^\circ$~\citep{10.1093/ptep/ptaa104, Esteban:2020cvm}. The luminosity distance in a standard flat $\Lambda \rm{CDM}$ cosmology is:
\begin{equation}
d_L(z) = (1+z) \frac{c}{H_0} \int_0^z \frac{dz^\prime}{\sqrt{\Omega_\Lambda + \Omega_M(1+z^\prime)^3}}\ ,
\end{equation}
where we use $H_0 =  67.4$~km~s$^{-1}$~Mpc$^{-1}$, $\Omega_M = 0.315$, and $\Omega_\Lambda = 0.685$~\citep{Planck:2018vyg, 10.1093/ptep/ptaa104}. 

The neutrino fluence at Earth  is
\begin{equation}
\Phi_{\nu_\alpha}(E_\nu) = \int_{t_i}^{t_f} dt\ F_{\nu_\alpha}(E_\nu, t) \ ,
\label{eq:fluence}
\end{equation}
where $F_{\nu_\alpha}(E_\nu, t)$ is given by Eq.~\ref{eq:neutrino_flux}, $t_i$ and $t_f$ are the onset and final times of neutrino production, respectively, measured by an observer at Earth. For the choked jet scenario, the integral in Eq.~\ref{eq:fluence} is  replaced by the product with the jet lifetime $t_j$. For CSM interaction, we fix the onset of our calculations $t_i \equiv t_{\rm{bo}} = (1+z) R_{\rm{bo}} / v_{\rm{sh}} $ and follow the neutrino signal up to $t_f \equiv t_{\rm{ext}}= (1+z) R_{\rm{ext}} /v_{\rm{sh}}$, where $ R_{\rm{ext}} = \min \left[ R_{\rm{CSM}}, R_{\rm{dec}} \right]$. In the last expression, $R_{\rm{dec}}$ is given by Eq.~\ref{eq:decRadius}. This choice is justified because  efficient particle acceleration takes place  for $ R \gtrsim R_{\rm{bo}}$ only; hence, no neutrinos can be produced before the breakout occurs. Second, for $R \gtrsim R_{\rm{ext}}$ either the CSM ends and there are no longer target protons for $pp$ interactions to occur, or the ejecta start to be decelerated and the neutrino signal quickly drops as $\propto v_{\rm{sh}}^2$~\citep{Petropoulou:2016zar}. Therefore, neutrino production is no longer efficient.

Both the cocoon model and the merger model predict the presence of  slow ejecta,  with $v_{\rm{s}} \simeq 0.01 c$. Nevertheless, the fast component of the ejecta in the cocoon model  sweeps up the CSM around the star; therefore, when the slow component emerges, there are no longer target protons for efficient $pp$ interactions to occur (in the assumption of spherical symmetry). As for the merger model, the slow outflow propagates into a highly dense and compact CSM. However, shocks in the equatorial region are radiative, and neutrinos should be produced with a maximum energy lower than the one of neutrinos produced in the fast outflow--CSM interaction (see e.g.~\cite{Fang:2020bkm}). Furthermore, the equatorial CSM has a  smaller extension than the polar one, and the corresponding neutrino production would last for a shorter time. As a consequence, we consider the neutrino signal from the fast outflow only.

\section{Neutrino signal from nearby sources}\label{sec:results}
In this section, we present our forecasts for the neutrino signal for the  the choked and CSM interaction models. We also discuss the number of neutrinos expected at the  IceCube Neutrino Observatory as well as the detection perspectives at  upcoming  neutrino detectors, such as IceCube-Gen2.

\subsection{Neutrino fluence}
\begin{figure*}
\centering
	\includegraphics[width=0.48\textwidth]{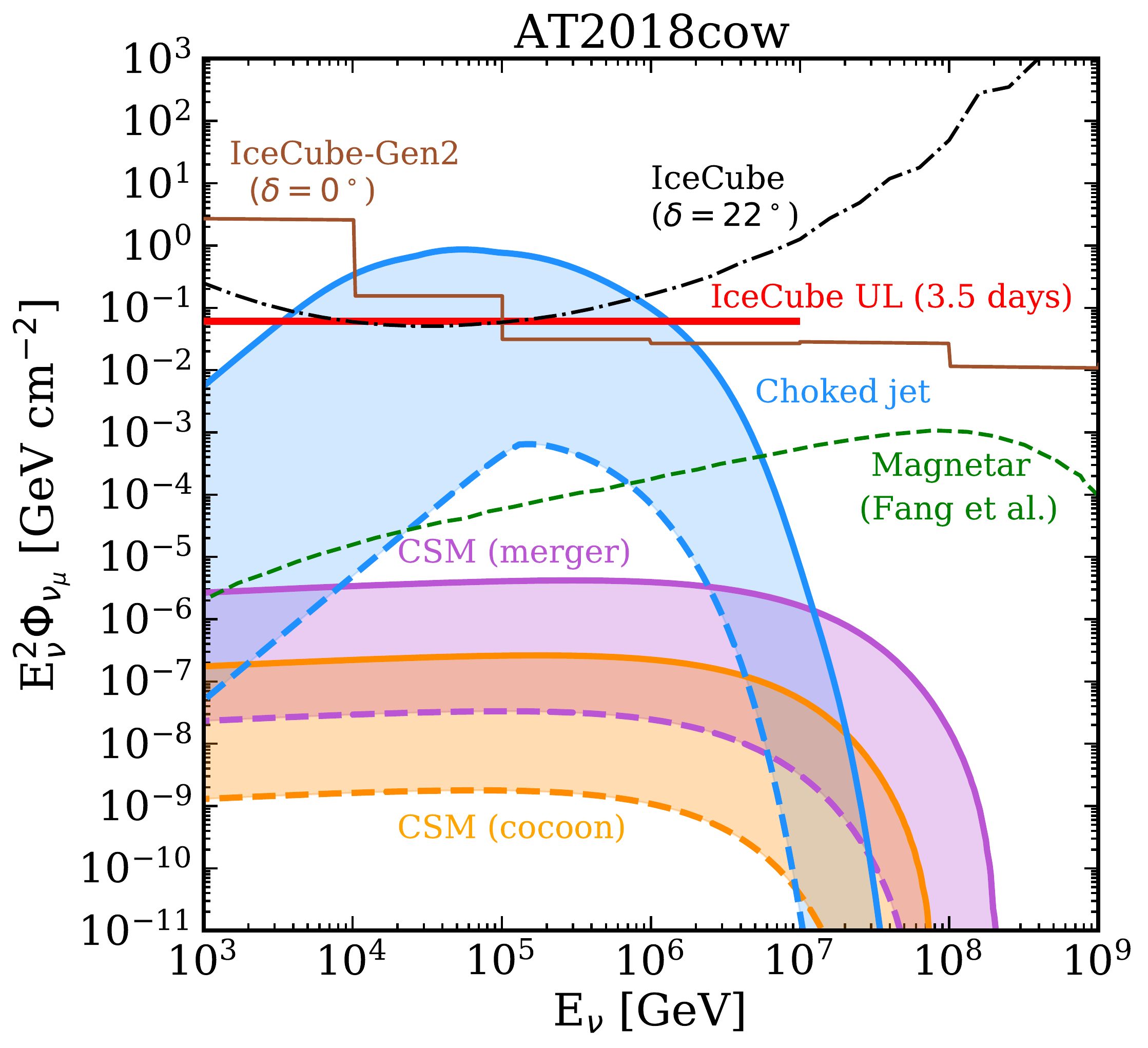}
	\includegraphics[width=0.48\textwidth]{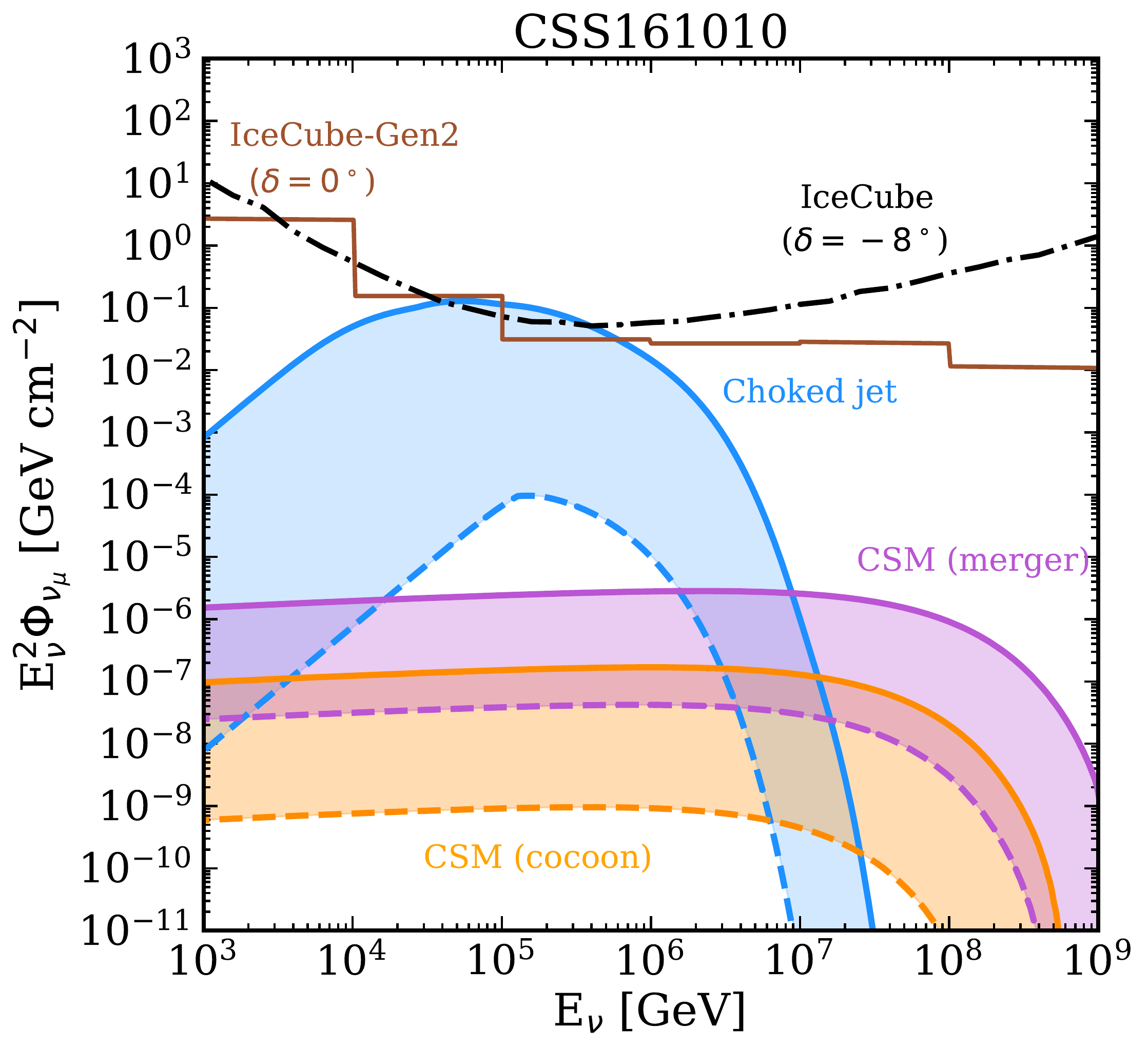}
	\caption{Muon neutrino fluence expected from AT2018cow (left panel, $z = 0.0141, \delta = 22^\circ$) and CSS161010 (right panel, $z = 0.034, \delta = -8^\circ$). The blue shaded region corresponds to the contribution to the neutrino signal from the choked jet, while the orange (purple) shaded region displays the signal from interaction between the CSM and the fast component of the outflow in the cocoon (merger) model. The continuous (dashed) lines are the upper (lower) limits on the neutrino fluence, corresponding to the ranges of parameter values listed in Table~\ref{table:parametersA}. The neutrino emission from the choked jet scenario is strongly dependent on the direction, while the one from the CSM scenarios is quasi-isotropic. The neutrino fluence in the choked jet scenario is shown in the most optimistic case of a jet oriented along the line of sight of the observer. 
For comparison, we show the results of~\cite{Fang:2018hjp} (green dashed line), corresponding to the neutrino fluence in the event that a magnetar powers AT2018cow. The sensitivity of IceCube for point sources is plotted at a declination $\delta = 22^{\circ}$ and $\delta = -8^{\circ}$~\citep{IceCube:2014vjc} (black dot-dashed lines), as measured for AT2018cow and CSS161010, respectively. The sensitivity of IceCube-Gen2 to a point source at $\delta = 0^\circ$ is also shown (sienna line). The neutrino fluence from the choked jet harbored in LFBOTs---if the jet points towards the observer---is comparable  with the sensitivities of  IceCube and IceCube-Gen2. For AT2018cow, we show the upper limit  set by IceCube on the time-integrated $\nu_\mu$ fluence (IceCube UL, red line), corresponding to the observation of two neutrino events in coincidence with AT2018cow~\citep{2018ATel11785....1B, Stein:2019ivm}.} 
	\label{fig:neutrinos}
\end{figure*}
For the choked jet scenario (see Sec.~\ref{sec:chokedjet}), for fixed isotropic equivalent energy $\tilde{E}_{j}^{\rm{iso}}$, we consider an envelope containing the expected neutrino fluence for  the allowed $(\tilde{L}_{j}, \tilde{t}_j)$ pairs.
As for CSM interaction in the cocoon model, we fix $R_{\rm{bo}} = R_{\rm{env}} = 3 \times 10^{13}$~cm up to $R_{\rm{ext}}$. Indeed, in the hypothesis of an extended stellar envelope surrounding the core of the star, the CSM is already optically thin at the edge of the envelope and radiation can escape as soon as the cocoon breaks out. As already pointed out, in the merger model the breakout radius is calculated by using Eq.~\ref{eq:breakoutRad} and it does not occur  too deep in the CSM, since the latter is not very dense. 

Figure~\ref{fig:neutrinos} shows the muon neutrino fluence expected from AT2018cow and CSS161010. The blue band corresponds to the neutrino fluence  from the choked jet, while the orange and purple bands represent the neutrino signal from CSM interaction in the cocoon and merger models, respectively. Each  band  reflects the uncertainties on the model parameters discussed in Sec.~\ref{sec:parameters} (see Table~\ref{table:parametersA}). The neutrino fluence for the  choked jet scenario is displayed for the optimistic case of a jet observed on axis.  If the jet axis should be perpendicular with respect to the line of sight of the observer,  no neutrino is expected.  In the following, we assume that the choked jet points towards the observer; this might have been the case for AT2018cow, since two neutrinos have been detected at IceCube in its direction~\citep{2018ATel11785....1B, Stein:2019ivm}---see discussion below.  On the other hand,  the emission from CSM interaction is approximately isotropic and hence observable from any viewing angle. This is  consistent with electromagnetic observations of LFBOTs: if a choked jet is harbored, no electromagnetic emission is expected.  The optical radiation is powered from the cooling of the cocoon, while the radio emission comes from the interaction of the cocoon with the CSM~\citep{Gottlieb:2022old}. In the merger model, the fast outflow  responsible for the high-energy neutrino emission   likely covers about $\gtrsim 70\%$ of the solid angle $4 \pi$~\citep{Metzger:2022xep}; hence, its emission is quasi-isotropic and visible from along any observer direction. 

Both for the cocoon  and  merger models, CSM interaction produce a smaller neutrino fluence than in the case of the choked jet model. Nevertheless, the merger model allows for a larger neutrino fluence compared to the cocoon one. This result is justified in the light of the larger CSM densities. Even though the stellar mass-loss rates are comparable, the wind speed is lower in the merger model than in the cocoon model ($10$~km s$^{-1}$ and $1000$~km s$^{-1}$, respectively; in the former model, it is generated by mass loss from the disk, while  it is due to mass loss from the progenitor star prior to its explosion in the latter model). If a choked jet is harbored in LFBOTs and points towards the observer, then it dominates the neutrino emission. The neutrino emission from  the choked jet model is in qualitative agreement with~\cite{Murase:2013ffa,He:2018lwb,Senno:2015tsn}, which focused on forecasting the neutrino production in gamma-ray bursts instead. 
Our results concerning the neutrino signal from ejecta-CSM interaction are valid for every model invoking the emission of a fast outflow propagating outwards in the CSM. On the contrary, neutrino emission from the choked jet is  model dependent. Recent numerical simulations show that efficient acceleration in jets can  occur if the jet is weakly or mildly magnetized~\citep{Gottlieb:2021pzr}; if this should be the case for LFBOTs, a dedicated investigation of the neutrino production in this scenario would be required. Furthermore, we have calculated the neutrino signal from the jet assuming that it is choked in the extended stellar envelope. As discussed in Sec.~\ref{sec:parameters} and shown in Fig.~\ref{fig:parameterSpace}, a choked jet may be harbored only  for certain pairs  of the jet luminosity and lifetime.

For comparison, in Fig.~\ref{fig:neutrinos},  we show the sensitivity of IceCube for point sources at the declination $\delta = 22^{\circ}$ (for AT2018cow) and $\delta = -8^{\circ}$ (for CSS161010)~\citep{IceCube:2014vjc} and  the projected sensitivity of IceCube-Gen2 for a point-like source at $\delta = 0^\circ$~\citep{IceCube-Gen2:2020qha}. If a source similar to AT2018cow (or CSS161010) were to be observed in the future at this declination by IceCube-Gen2, the detection chances of neutrinos from the choked jet scenario would be comparable to the ones of IceCube. This is mainly due to the fact that the sensitivity of IceCube-Gen2 will be better than the one of IceCube in the PeV--EeV energy range but comparable at lower energies;  the fluence from the choked jet peaks in the TeV--PeV range. As for the CSM interaction, the neutrino fluence lies well below the sensitivity curve of both IceCube and IceCube-Gen2. 

Other neutrino detectors are planned to be operative in the future, such as GRAND 200k~\citep{GRAND:2018iaj}, RNO--G~\citep{RNO-G:2020rmc} and POEMMA~\citep{POEMMA:2020ykm}. These neutrino telescopes aim to probe ultra high energy neutrinos, but their sensitivity in the PeV--EeV energy range is worse than the one of IceCube-Gen2; therefore we do not show them in Fig.~\ref{fig:neutrinos}.

\begin{figure}
\hspace{-1.0cm}
\includegraphics[width=0.55\textwidth]{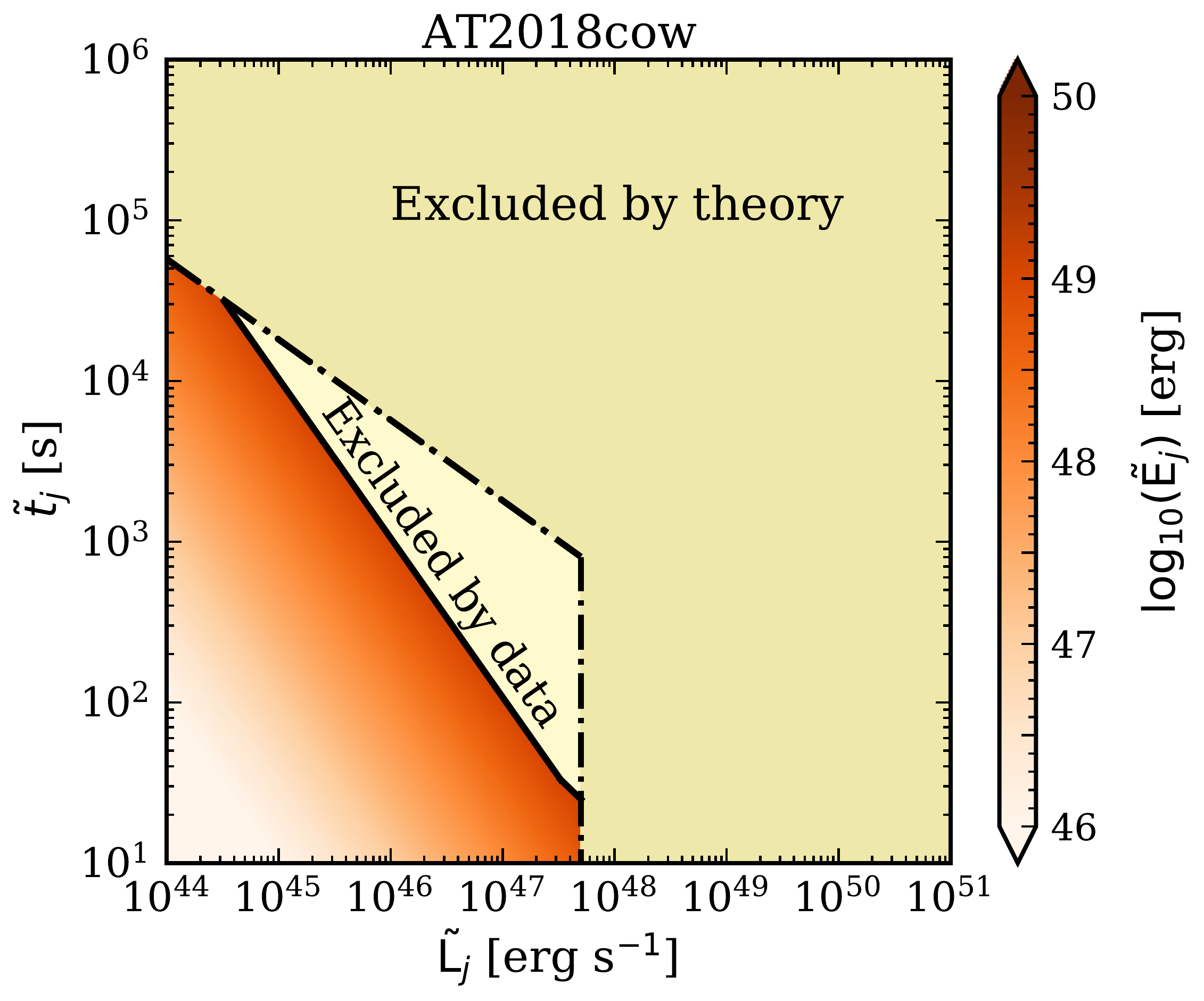}
\caption{Contour plot of the jet energy $\tilde{E}_{j}$ in the parameter space spanned by  $(\tilde{L}_{j}, \tilde{t}_j)$ for AT2018cow and $\Gamma = 100$. Part of the  parameter space allowed in Fig.~\ref{fig:parameterSpace}  is excluded by the IceCube neutrino data  (light yellow region), since the correspondent neutrino emission would overshoot the upper limit set by IceCube  on the time-integrated $\nu_\mu$ flux from for AT2018cow~\citep{2018ATel11785....1B, Stein:2019ivm}. Another portion of the parameter space (dark yellow region) is excluded by theoretical arguments, as already shown in Fig.~\ref{fig:parameterSpace}. For $\Gamma = 10$, the region of the parameter space excluded by the IceCube data is smaller and overlaps with the one excluded by theory. The region of the parameter space excluded by the IceCube data is obtained under the assumption of an on-axis  choked jet, see discussion in the main text.}
\label{fig:cowPS}
\end{figure}

In Fig.~\ref{fig:neutrinos},  we plot the upper limit set by IceCube on the muon neutrino fluence for AT2018cow. 
This upper limit corresponds to the observation of two  IceCube neutrino events in coincidence with AT2018cow at $1.8 \sigma$ confidence level within a time window of $3.5$~days  after the optical discovery~\citep{2018ATel11785....1B, Stein:2019ivm}. The envelope obtained for AT2018cow  overshoots this limit  for $\tilde{E}_{j}^{\rm{iso}} \gtrsim 10^{52}$~erg. Interestingly, $\tilde{E}_{j}^{\rm{iso}} \gtrsim 10^{52}$~erg falls in the range inferred by electromagnetic observations, see Table~\ref{table:parametersA}.  This finding intriguingly suggests that  existing neutrino data may further restrict the allowed parameter space shown in Fig.~\ref{fig:parameterSpace} for AT2018cow, as displayed in Fig.~\ref{fig:cowPS}. No neutrino search has been performed in the direction of CSS161010 instead. 

As discussed in Sec.~\ref{sec:parameters}, the CO of LFBOTs could be a magnetar. In this case, high-energy neutrinos could be produced in the proximity of the magnetar~\citep{Murase:2009pg, Fang:2013vla, Fang:2017tla}. Protons (or other heavier nuclei) may be accelerated in the magnetosphere and then interact with  photons and baryons in the ejecta shell surrounding the CO. Both $p \gamma$ and $pp$ interactions can efficiently produce neutrinos in the PeV--EeV energy band. The neutrino production from a millisecond magnetar has been investigated in~\cite{Fang:2020bkm} for AT2018cow. We show the expected muon neutrino fluence at Earth  obtained in~\cite{Fang:2020bkm} in Fig.~\ref{fig:neutrinos}  for comparison with the other scenarios explored in this paper. For CSS161010, we expect a neutrino fluence qualitatively similar to the one considered for AT2018cow.

If a magnetar is the central engine of LFBOTs, its contribution to the neutrino fluence would be relevant in the PeV--EeV band, at energies higher  than the typical ones for neutrino emission from the choked jet and CSM interaction. Note that the comparison between the fluence from the magnetar and our results is consistent as for   the energetics of the CO. Indeed, the  set of parameters adopted by~\cite{Fang:2018hjp} leads to  $\tilde{E} \simeq 10^{50}$--$10^{51}$~erg injected by the magnetar in its spin-down time, $\tilde{t}_{\rm{sd}} \simeq 8.4 \times 10^{3}$--$8.4 \times 10^4 $~s. If a jet is launched by the magnetar, then these quantities correspond to its energy and its lifetime, consistently with the ranges we are exploring in our work.  

The radio extension of IceCube-Gen2~\citep{IceCube-Gen2:2020qha}, as well as the neutrino facilities GRAND200k~\citep{GRAND:2018iaj}, POEMMA~\citep{POEMMA:2020ykm} and RNO-G~\citep{RNO-G:2020rmc} will be  more sensitive than IceCube~\citep{IceCube-Gen2:2020qha} for what concerns the emission of neutrinos in the magnetar scenario and they may detect neutrinos from sources similar to AT2018cow, occurring at a smaller distance.

\subsection{Neutrino event rate}\label{sec:eventRate}
\label{sec:event_rate}
\begin{figure*}[ht]
\centering
\includegraphics[width=0.48\textwidth]{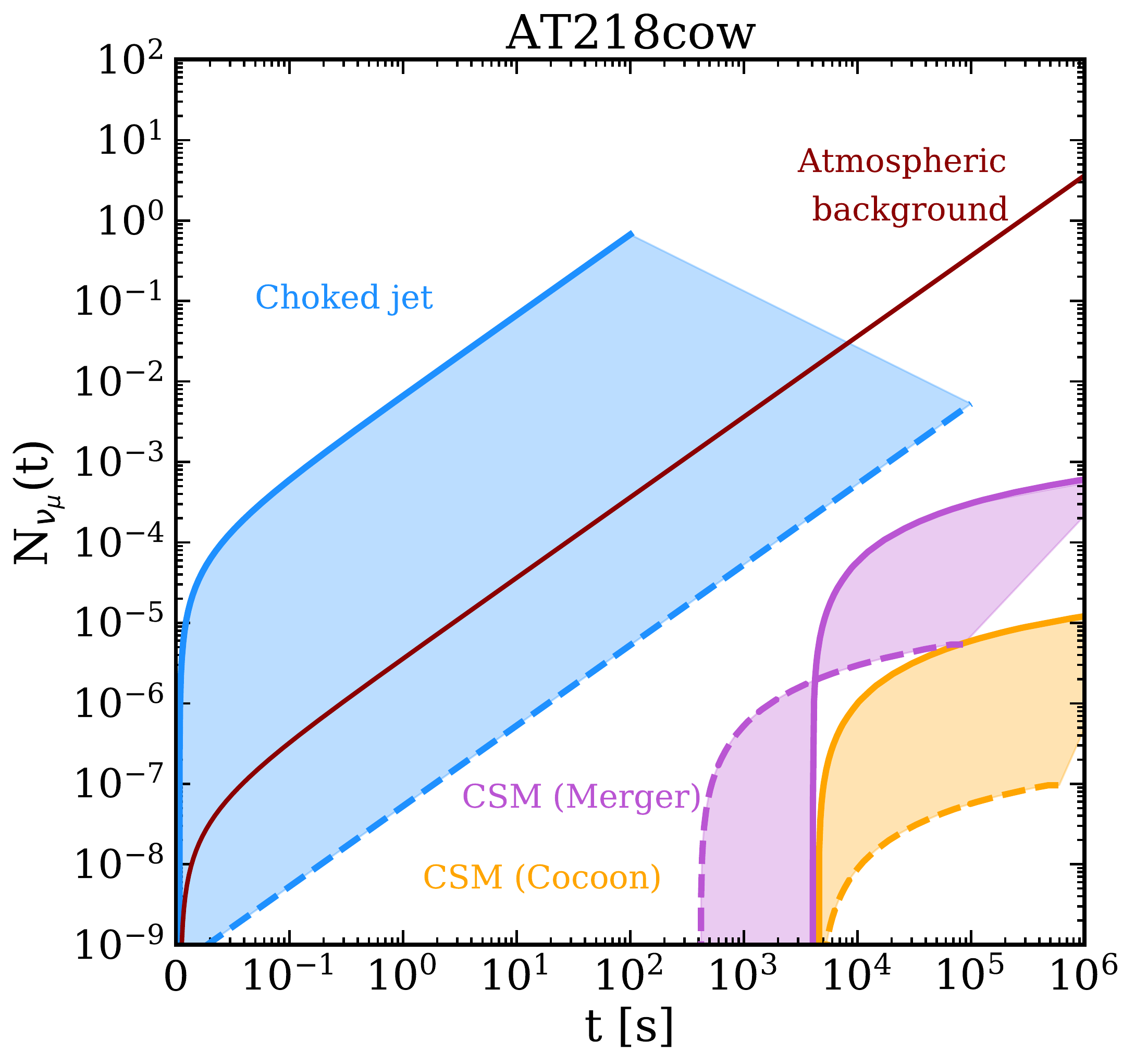}
\includegraphics[width=0.48\textwidth]{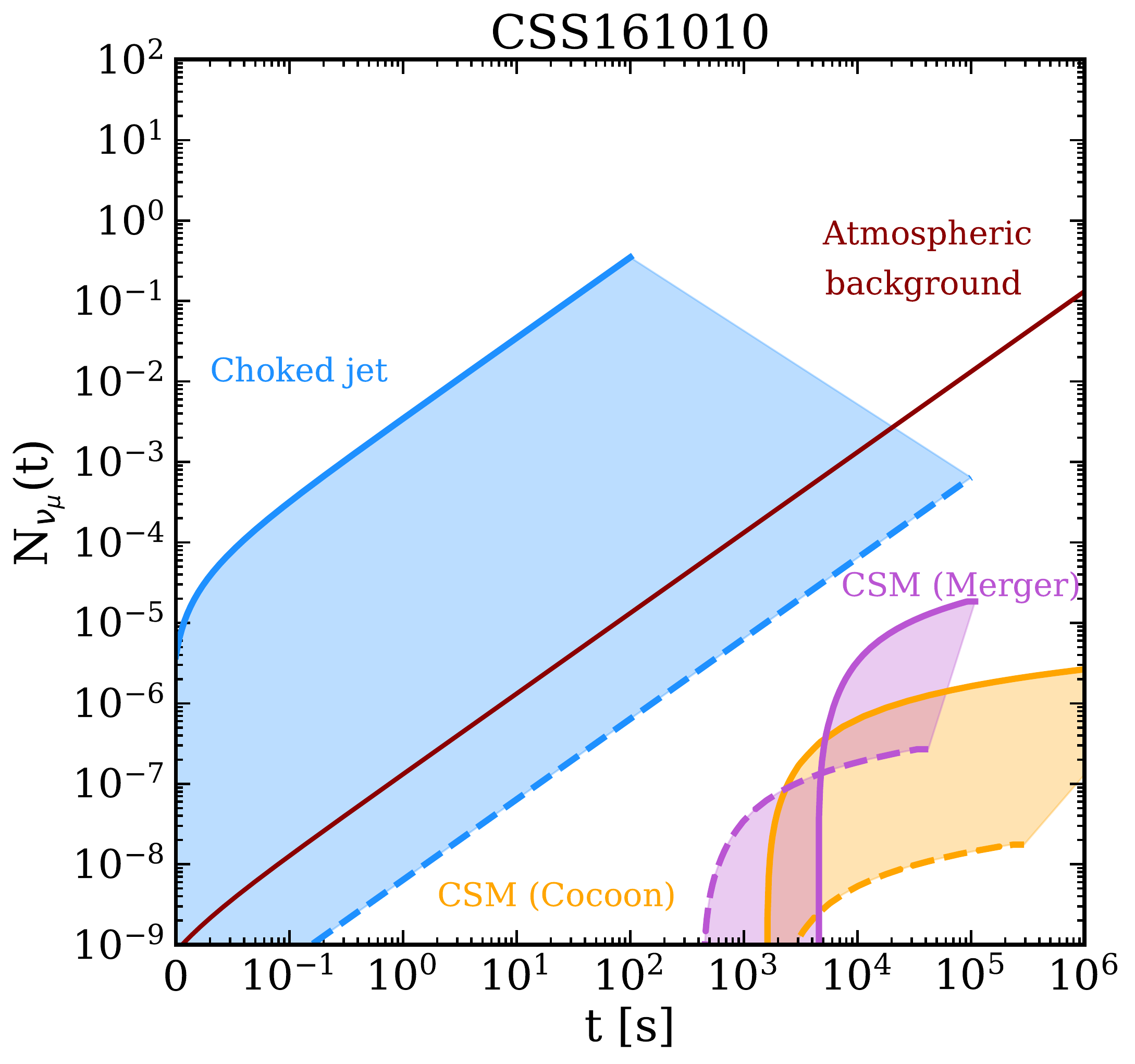}
\caption{Cumulative number of muon neutrinos  for AT2018cow (left panel) and CSS161010 (right panel) expected at the IceCube Neutrino Observatory; see Table~\ref{table:parametersA}. The blue shaded region corresponds to the contribution from the choked jet (when the latter is observed on-axis); the orange (purple) shaded region corresponds to neutrinos from CSM interaction in the cocoon (merger) model. The rate of neutrinos from the choked jet is expected to be constant in the approximation that $N$ internal shocks occur in the jet during its lifetime and that each of them produces the same neutrino signal. Therefore, the neutrino signal grows linearly with time up to the end of the jet lifetime. For CSM interaction, the number of neutrinos rapidly increases and  settles to a constant value since the proton injection is balanced by $pp$ energy losses.  The upper and lower limits of each band correspond to the same uncertainty ranges in Table~\ref{table:parametersA} and Fig.~\ref{fig:neutrinos}, except for the upper limit for the choked jet scenario in AT2018cow for which we take $\tilde{E}_{j}^{\rm{iso}} = 10^{51}$~erg, consistently with the IceCube constraints--see Fig.~\ref{fig:cowPS}. 
The brown line shows the cumulative number of atmospheric neutrinos (which constitutes a background for the detection of astrophysical neutrinos), which increases linearly with time.}
\label{fig:cumulative}
\end{figure*}
Given the muon neutrino fluence up to the time $t$, $\Phi_{\nu_\mu}(E_{\nu}, t)$, the cumulative number of muon neutrinos expected at IceCube up to the same time is
\begin{equation}
N_{\nu_\mu}(t) = \int_{E_{\nu, {\rm{min}}}}^{E_{\nu, {\rm{max}}}}dE_\nu \Phi_{\nu_\mu,}(E_{\nu}, t) A_{\rm{eff}}(E_\nu, \delta) \; ,
\label{eq:neutrinoN}
\end{equation}
where $E_{\nu, \rm{min}} = 10^2$~GeV and $E_{\nu, \rm{max}}  = 10^{10}$~GeV are the minimum and maximum neutrino energies, respectively, and $A_{\rm{eff}}(E_\nu, \delta)$ is the effective area  as a function of energy and for a fixed source declination $\delta$~\citep{IceCube:2021xar}.
The background of atmospheric muon neutrinos can be estimated following~\cite{Razzaque:2014ola}:
\begin{equation}
N_{\nu_\mu, \rm{back}}(t) = \pi \Delta \delta^2  \int_{E_{\nu, \rm{min}}}^{E_{\nu, {\rm{max}}}} d E_\nu  A_{\rm{eff}}(E_\nu, \delta) \Phi^{\rm{atm}}_{\nu_\mu}(E_{\nu}, \theta, t)\ ,
\label{eq:background}
\end{equation}
where $\Phi^{\rm{atm}}_{\nu_\mu}(E_{\nu}, \theta, t)$ is the fluence of atmospheric neutrinos  at the time $t$, from the zenith angle $\theta$ and $\Delta \delta \simeq 2.5^\circ$ is the width of the angular interval within which is defined the effective area $A_{\rm{eff}}(E_\nu, \delta)$ of IceCube. For IceCube, the  relation $\theta = 90^\circ + \delta$ holds~\citep{IceCube:2016tpw}. 
We compute the atmospheric background by using the model presented in~\cite{stanev2010high,Gaisser:2002jj, Gaisser:2019efm}.

We show the cumulative number of neutrinos from the choked jet scenario (for a jet pointing towards the observer) and CSM interaction (both for the cocoon and  merger models) as  functions of time both for AT2018cow and CSS161010 in Fig.~\ref{fig:cumulative}. Note that for AT2018cow,  the upper limit of the choked jet scenario is calculated by assuming  $\tilde{E}_{j}^{\rm{iso}}= 10^{51}$~erg, in agreement with the allowed region of the parameter space shown in  Fig.~\ref{fig:cowPS}. The upper and lower limits for the  cumulative number of neutrinos in the CSM interaction models for AT2018cow  and for all the scenarios considered for CSS161010 are the same as the ones in Table~\ref{table:parametersA}. The thick lines denote the duration of the signal, which can last up to  a few months for CSM interaction. As for the choked jet, the neutrino rate is expected to be constant during the jet lifetime, in the simple approximation that $N$ internal shocks occur in the jet during this period and  each of them produces the same neutrino signal. Hence, the cumulative neutrino rate from the choked jet grows linearly with time up to the jet lifetime. For CSM interaction, the number of neutrinos rapidly increases after the breakout and then reaches a plateau since the proton injection is balanced by $pp$ energy losses. The atmospheric background neutrinos increase linearly with time. The background  is expected to dominate over the signal from CSM interaction, both for the cocoon and merger models; on the contrary, the background becomes  comparable to the choked jet signal at times larger than the jet lifetime. 

\subsection{Detection prospects for AT2018cow and CSS161010}
\label{sec:detection_present}
The neutrino signal from LFBOTs overlaps in energy with the atmospheric neutrino background. In order to gauge the possibility of discriminating the LFBOT signal from the one of atmospheric neutrinos, we compare the total number of muon neutrinos of astrophysical origin $N_{\nu_\mu, \rm{astro}}$ with the total number of background atmospheric  neutrinos $N_{\nu_\mu, \rm{back}}$. 
The former is given by the sum between contributions from the choked jet and CSM interaction  in the cocoon model and by CSM interaction only in the merger model. Each contribution is computed by relying on Eq.~\ref{eq:neutrinoN} and integrating over the  duration of the neutrino production, defined for each case in Sec.~\ref{sec:fluxEarth}. The latter is obtained through Eq.~\ref{eq:background}, during the duration of neutrino production for each model. 

Below $100$~TeV, the astrophysical neutrino events need to be carefully discriminated against the atmospheric ones. Hence,  we consider two scenarios: a conservative energy cutoff in Eq.~\ref{eq:neutrinoN}, $E_{\nu, \min} = 100$~TeV (corresponding to the case when atmospheric neutrino events cannot be distinguished from the astrophysical ones below $100$~TeV) and a low energy cutoff, $E_{\nu, \min} = 100$~GeV (representative of the instance of full discrimination of the events of astrophysical origin). 

Our results are summarized in Table~\ref{table:signalness}.
The number of astrophysical neutrinos expected in   the cocoon model is larger than the  number of atmospheric neutrinos, both for AT2018cow and CSS161010, when the energy cutoff $E_{\nu, \min} = 100$~TeV is adopted. Hence, the detection chances of astrophysical neutrinos above $100$~TeV may be promising, if a choked jet pointing towards the observer  is harbored in LFBOTs.
The number of astrophysical neutrinos may  instead be smaller than or comparable to the atmospheric background for the merger model, therefore, the background signal cannot be fully  discriminated; this is especially evident for $E_{\nu, \min} = 100$~GeV. 

\begin{table*}
\centering
\caption{Total number of astrophysical neutrinos ($N_{\nu_\mu, \rm{astro}}$) and atmospheric  neutrinos ($N_{\nu_\mu, \rm{back}}$) in the cocoon (including choked jet and CSM interaction) and merger models, for $E_{\nu, \min} = 100$~TeV. The correspondent neutrino numbers obtained by adopting  $E_{\nu, \min} = 100$~GeV are displayed in parenthesis. The range of variability corresponds to the upper and lower limits shown in Fig.~\ref{fig:cumulative}.}
\begin{tabularx}{.75\textwidth}{ccc}
\toprule
\toprule
${N_{\nu_\mu}}$ &  {AT2018cow} &  {CSS161010}\\	
\toprule
 & Cocoon model  &  \\
\toprule
${N_{\nu_\mu, \rm{astro}}}$   &$3 \times 10^{-3}$--$0.15 \; (7 \times 10^{-3} - 0.67)$ &   $3 \times 10^{-4}$--$0.23 \; (4 \times 10^{-4} - 0.35)$ \\
${N_{\nu_\mu, \rm{back}}}$ & $9 \times 10^{-4}-3 \times 10^{-3} \; (2.23-9.71)$ &  $5 \times 10^{-4} - 1.4 \times 10^{-2} \; (2.6 \times 10^{-2} - 0.64)$ \\

\toprule
 & Merger model &  \\
\toprule
${N_{\nu_\mu, \rm{astro}}}$   & $1.5 \times 10^{-6}$--$2.1 \times 10^{-4} \; (1.5 \times 10^{-6}- 2 \times 10^{-4})$      &$6.5 \times 10^{-7}$--$4.5 \times 10^{-5} \; (8 \times 10^{-7} - 5 \times 10^{-5})$ \\
${N_{\nu_\mu, \rm{back}}}$ & $1 \times 10^{-4}-3 \times 10^{-3} \; (0.32-8)$ &  $ 8 \times 10^{-5} - 2 \times 10^{-3} \; (3.7 \times 10^{-3} - 9.2 \times 10^{-2})$ \\
\toprule
\toprule
\end{tabularx}
\label{table:signalness} 
\end{table*}

In the event of detection of one or a few neutrinos from LFBOTs and depending on the number of undetected sources from the LFBOT population, the actual neutrino flux could be smaller than the one  estimated by relying on the detected events. For this reason, we need to correct for  
the Eddington bias on neutrino observations~\citep{Strotjohann:2018ufz}. 
Assuming  that the local rate of LFBOTs is $\sim 0.4 \%$ of the core-collapse SN rate~\citep{Coppejans:2020nxp}, we consider the 
effective density integrated over the cosmic history of LFBOTs to be $\mathcal{O}(10^4)$~Mpc$^{-3}$. The latter  has been computed by assuming the  density  of core-collapse SNe equal to $1.07 \times 10^7$~Mpc$^{-3}$~\citep{Yuksel:2008cu,Vitagliano:2019yzm} and   the redshift evolution of  LFBOTs  identical to the one of the star formation rate. After aking into account these inputs, from Fig.~2 of~\cite{Strotjohann:2018ufz}, 
we find that the number of expected events  in Table~\ref{table:signalness} could be compatible with the observation of $1$--$3$ neutrino events both from AT2018cow and CSS161010.

The IceCube Neutrino Observatory reported the detection of two track-like neutrino events 
in the direction of AT2018cow compatible with the expected number of atmospheric neutrino events~\citep{2018ATel11785....1B}. Our findings hint that the observation of two neutrino events may also be  compatible with the expected number of neutrinos of astrophysical origin. 
A dedicated neutrino search  in the direction of CSS161010,  during the time when the transient was electromagnetically bright, would be desiderable.

\subsection{Future detection prospects}
\begin{figure}
\centering
\includegraphics[width=0.48\textwidth]{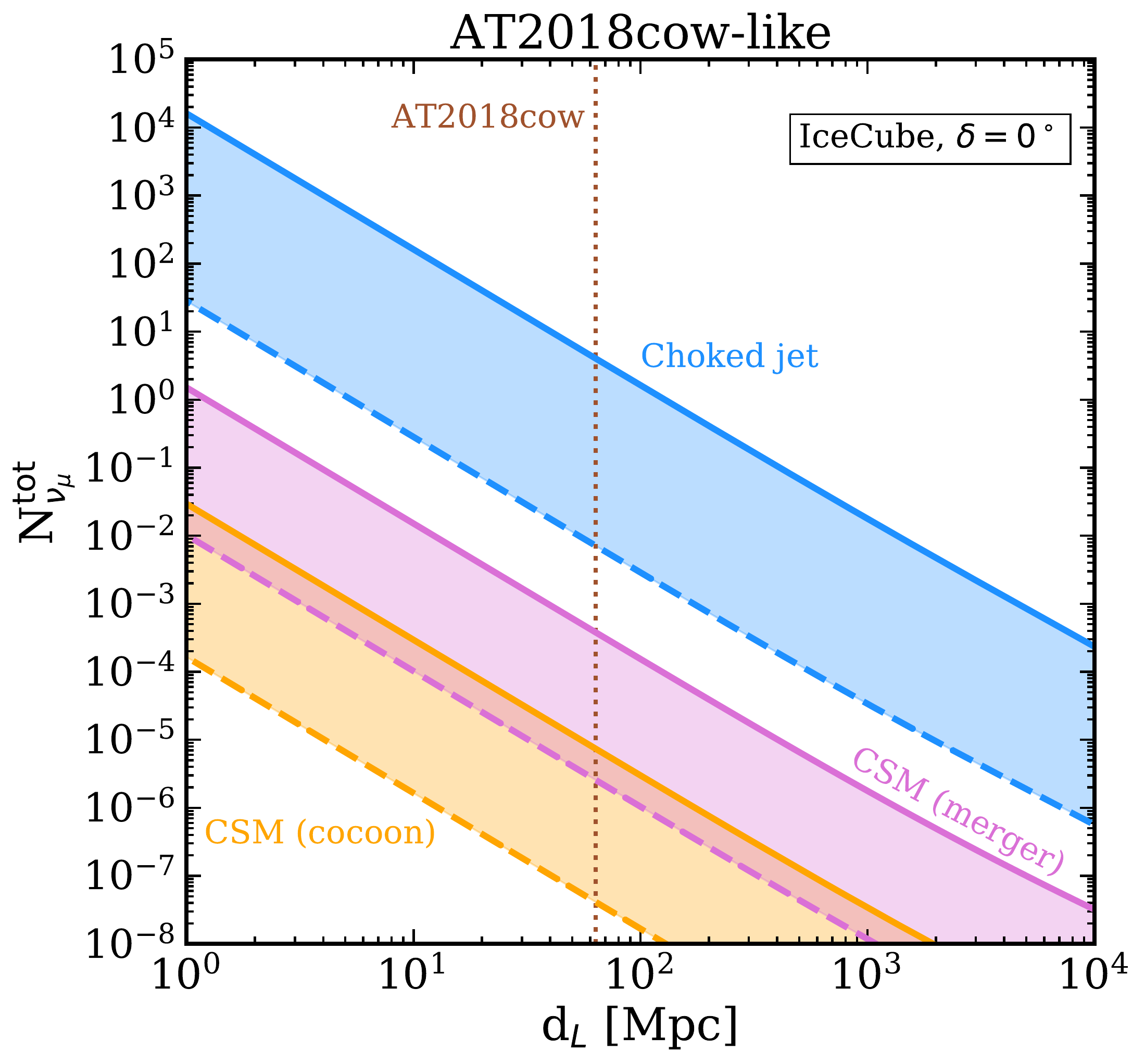}
\caption{Total number of muon neutrinos expected at the IceCube Neutrino Observatory as a function of the luminosity distance for an AT2018cow-like source from the choked jet pointing towards the observer (blue shaded region) and CSM interaction in the cocoon and merger models (orange and purple shaded regions, respectively). The bands are obtained by adopting the parameter uncertainty ranges  listed in Table~\ref{table:parametersA} for AT2018cow. The source is placed at $\delta = 0^\circ$. The brown vertical line marks the distance of AT2018cow to guide the eye. The number of neutrinos decreases as a function of the luminosity distance, as expected. }
\label{fig:neutrinoVSdistance}
\end{figure}

The rate of LFBOTs and its redshift dependence are still very uncertain. In oder to forecast the detection prospects in neutrinos for upcoming LFBOTs, we consider an LFBOT with properties similar to the ones of  AT2018cow (see Table~\ref{table:parametersA}).
Figure~\ref{fig:neutrinoVSdistance} shows the  total number of neutrinos expected at the IceCube Neutrino Observatory in the choked jet scenario and for CSM interaction (both in the cocoon and merger models) as  functions of the luminosity distance of the AT2018cow-like source;  of course, similar conclusions would hold  for an LFBOT resembling the CSS161010 source, see Figs.~\ref{fig:neutrinos} and \ref{fig:cumulative}. We assume  the upper and lower limits for AT2018cow listed in Table~\ref{table:parametersA}, since  the neutrino constraints shown in Fig.~\ref{fig:cowPS} do not hold for this source. We assume that  the source is at $\delta = 0^\circ$, in order to guarantee the maximal effective area at IceCube~\citep{IceCube:2021xar}, and  perform the integral in Eq.~\ref{eq:neutrinoN} between the initial ($t_i$) and final ($t_f$) times of neutrino production as described in Sec.~\ref{sec:fluxEarth}. Furthermore, we adopt the conservative lower energy cutoff $E_{\nu, \min} = 100$~TeV, in order to better discriminate neutrinos of astrophysical origin from atmospheric background neutrinos.

Figure~\ref{fig:neutrinoVSdistance} shows that the number of neutrinos expected in the choked jet scenario for an AT2018cow-like source located at $1$~Mpc $\lesssim d_L \lesssim 10^4$~Mpc is  $10^{-6} \lesssim N_{\nu_\mu}^{\rm{tot}} \lesssim 10^4$ if the jet  points towards the observer.
As for CSM interaction, the number of expected neutrinos for the same source located at $1$~Mpc $\lesssim d_L \lesssim 10^4$~Mpc is  for the cocoon model (merger model) is  $2 \times 10^{-12} \lesssim N_{\nu_\mu}^{\rm{tot}} \lesssim 3 \times 10^{-2}$ ($ 10^{-10} \lesssim N_{\nu_\mu}^{\rm{tot}} \lesssim 2$). 
 We  expect comparable or better detection chances for IceCube-Gen2 (see  Fig.~\ref{fig:neutrinos}).
We stress that a detailed statistical analysis may provide improved detection prospects, but  this is out of the scope of this work. Nevertheless, our results are an intriguing guideline for upcoming follow-up  neutrino searches of LFBOTs.

\section{Diffuse neutrino emission}\label{sec:diffuse}
\begin{figure}
\centering
\includegraphics[width=0.48\textwidth]{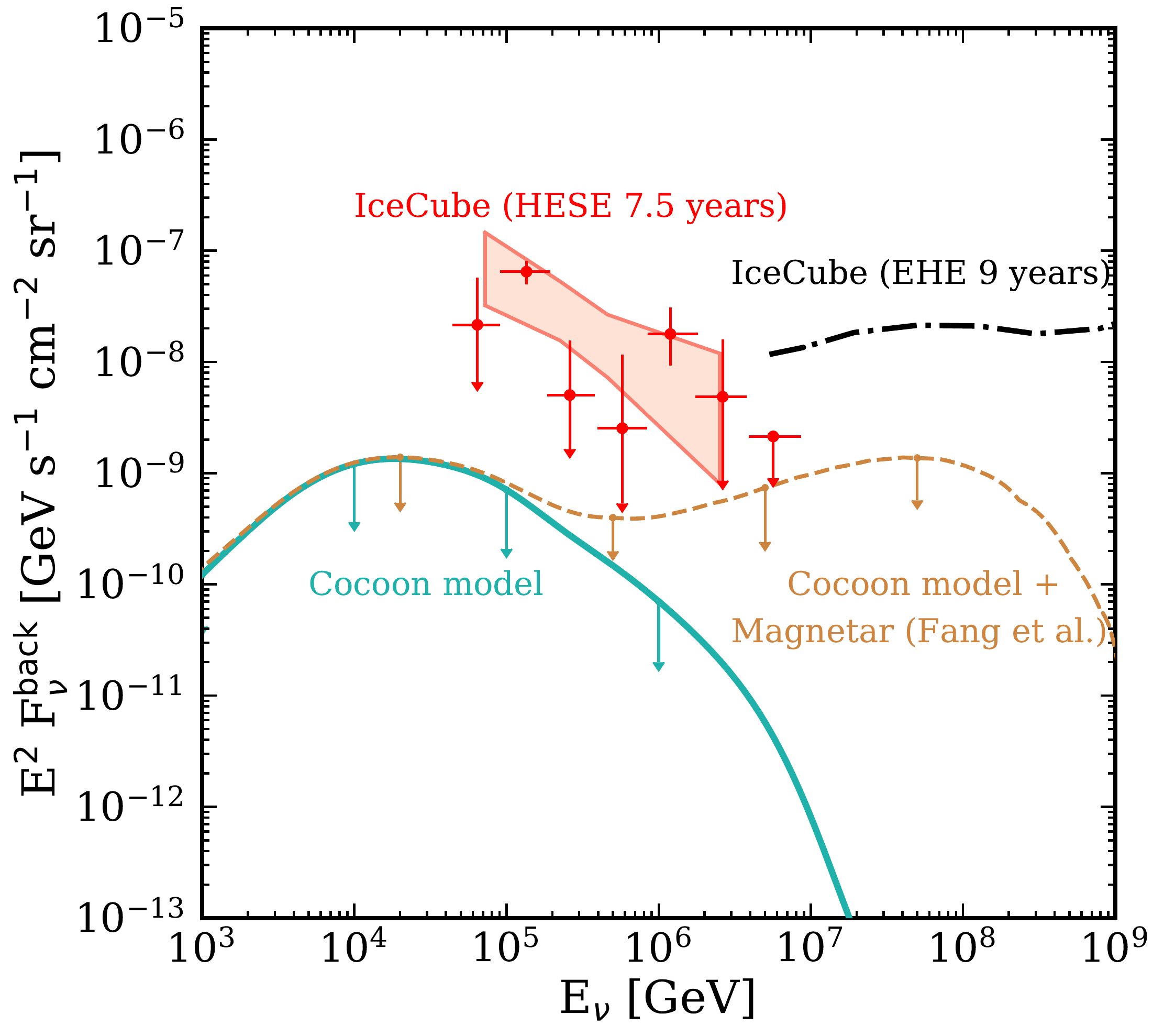}
\caption{Upper limit on the all-flavor diffuse neutrino flux from LFBOTs  obtained by including the contribution from a choked jet and CSM interaction (cocoon model; seagreen solid line) as a function of the neutrino energy. We show the cocoon model only, since it includes both the choked jet and CSM interaction; the merger model would give rise to a diffuse flux lying well below the seagreen line. For comparison, the upper limit obtained including both the cocoon model and the contribution from the magnetar (taken from~\cite{Fang:2018hjp}) is also shown (light-brown dashed line). 
The pink band corresponds to the fit to the $7.5$ year IceCube high energy starting events (HESE), plotted as red datapoints~\citep{IceCube:2020wum}. The black dot-dashed line corresponds to the $9$~year extreme-high-energy (EHE) $90 \%$ upper limit set by the IceCube Neutrino Observatory~\citep{IceCube:2018fhm}. The diffuse neutrino background from LFBOTs lies below IceCube data.}
\label{fig:diffuse}
\end{figure}
Despite the growing number of neutrino events routinely detected at IceCube, the origin of the observed diffuse neutrino background is still unknown. Several source classes have been proposed as major contributors to the observed diffuse flux, such as gamma-ray bursts, cluster of galaxies, star-forming galaxies, tidal distruption events, and SNe~\citep{Meszaros:2017fcs, Ahlers:2018fkn, Vitagliano:2019yzm, Meszaros:2015krr,Pitik:2021xhb,Murase:2015ndr, Waxman:2015ues, Tamborra:2014xia, Zandanel:2014pva, Wang:2015mmh, Dai:2016gtz, Senno:2016bso,Lunardini:2016xwi, Sarmah:2022vra}.  As discussed in Sec.~\ref{sec:results}, LFBOTs have favorable detection chances in neutrinos, hence we now explore the  contribution of LFBOTs to the diffuse neutrino background. 

The diffuse neutrino background is 
\begin{equation}
    F^{\rm{back}}_\nu = \frac{c}{4 \pi H_0} \int_{0}^{z_{\rm{max}}} dz \frac{f_b R_{\rm{SFR}}(z)}{\sqrt{\Omega_M(1+z)^3 + \Omega_\Lambda}} \phi_{\nu}(E^\prime_\nu)\ ,
\end{equation}
where $z_{\rm{max}} = 8$, $\phi_{\nu}(E^\prime_\nu)$ is the differential neutrino number from a single burst (in units of GeV$^{-1}$; defined multiplying Eq.~\ref{eq:fluence} by the luminosity distance), $E^\prime_\nu = E_\nu (1+z) / \Gamma$ (where $\Gamma = 1$ for CSM interaction). The beaming factor is given by $f_b = \Omega / 4 \pi \simeq \theta_j^2 /2$ for the choked jet, while $f_b = 1$ for CSM interaction. The factor takes into account the beaming of the jet within an opening angle $\theta_j$. The beaming is not relevant in $pp$ interactions, since they originate from the cocoon (or the polar fast outflow) whose opening angle is wider than the one of the jet~\citep{Gottlieb:2021srg}. Therefore, the geometry of the outflow can be assumed to be spherical. 

So far, the luminosity function for LFBOTs is not available because only a few transients have been identified as belonging to  this emerging class. Thus,  we fix the isotropic equivalent energy of the choked jet $\tilde{E}_{j}^{\rm{iso}} = 10^{51} \; \rm{erg}$, its Lorentz factor $\Gamma = 100$, and assume that it is  representative of the whole population. For computing the contribution to the diffuse neutrino background from CSM interaction, we assume $M_{\rm{ej}} = 10^{-2} M_\odot$, $\dot{M}= 10^{-3} M_{\odot} \; \rm{yr}^{-1}$, $v_{\rm{w}} = 1000$~km s$^{-1}$, and $v_{\rm{sh}} = 0.3 c$ as representative values. 

We assume that the redshift evolution of LFBOTs follows the star formation rate, $R_{\rm{SFR}}(z)$~\citep{Yuksel:2008cu}: 
\begin{equation}
    R_{\rm{SFR}}(z)= R_{\rm{FBOT}}(z = 0) \biggl[ (1+z)^{-34}+ \biggr( \frac{1+z}{5000} \biggl)^{-3}+ \biggl( \frac{1+z}{9}\biggr)^{-35} \biggr]
\end{equation}
where the local rate of LFBOTs is assumed to be $R_{\rm{FBOT}} (z = 0) \lesssim 300 $ Gpc$^{-3}$ yr$^{-1}$~\citep{Coppejans:2020nxp,Ho:2021fyb}.  

Figure~\ref{fig:diffuse} shows the  upper limit to the diffuse neutrino background resulting from the choked jet and CSM interaction  (cocoon model; seagreen solid line). For comparison, we also show the upper limit on the total diffuse emission when we include the contribution from a millisecond magnetar, i.e.~when we sum up the diffuse emission from choked jet, CSM interaction and the magnetar itself (light-brown dashed line). The diffuse emission from the magnetar only has been taken from~\cite{Fang:2018hjp} and it has been rescaled to the local rate assumed in this paper, referred to the subclass of LFBOTs.
Note that here we consider the cocoon model only, since it includes both a choked jet and CSM interaction and thus it would lead to the most  optimistic estimation of the expected neutrino background.  If the merger model is adopted,  the resulting diffuse neutrino background is flat at low energies, with an energy cutoff around $10^7$~GeV; hence, the merger model would give rise to a diffuse emission below the seagreen line in Fig.~\ref{fig:diffuse} and consistent with the upper limit we are showing. 

We compare our results with the flux constraints from the $7.5$~year high-energy starting event data set (HESE 7.5yr)~\citep{IceCube:2020wum} and the $9$ year extreme-high-energy (EHE) $90 \%$ upper limit set by the IceCube Neutrino Observatory~\citep{IceCube:2018fhm} in Fig.~\ref{fig:diffuse}. Our results suggest that LFBOTs do not constitute the bulk of the diffuse neutrino flux detected by the IceCube Neutrino Observatory. Nevertheless, typical energies of these objects might be larger that the ones assumed in this work, resulting in a larger diffuse neutrino emission.

\section{Conclusions}\label{sec:conclusions}
Despite the growing number of observations of LFBOTs, their nature remains elusive. Multi-messenger observations could be crucial to gain insight on the source engine.

In this paper, we consider the  scenarios proposed in \cite{Gottlieb:2022old} (cocoon model) and \cite{Metzger:2022xep} (merger model) for powering LFBOTs and aiming to fit  multi-wavelength electromagnetic observations and  mounting evidence for asymmetric LFBOT outflows. In the cocoon model, neutrinos could be produced in the  jet choked  within the extended envelope of the collapsing massive star. The existence of a jet harbored in LFOBTs is highly uncertain, and certain conditions on its luminosity and lifetime must be satisfied for the jet to be choked. If a jet is launched by the CO and choked, it contributes to inflate the cocoon, the latter breaks out of the stellar envelope and interacts with the CSM;  neutrinos could also be produced at the collisionless shocks occurring at the interface between the cocoon and the CSM. In the merger model, a black hole surrounded by an accretion disk forms as a result of the merger of a Wolf-Rayet star with a black hole. The disk outflow in the polar region propagates in the CSM, possibly giving rise to neutrino production. 

By using the model parameters inferred from the electromagnetic observations of two among the most studied LFBOTs, AT2018cow and CSS161010, we find that neutrinos with energies up to $\mathcal{O}(10^9)$~GeV could be produced  in the cocoon and  merger models. The neutrino signal from the choked jet would be detectable only if the observer line of sight is located within the opening angle of the jet. If this is the case, the upper limit on the neutrino emission set by the IceCube Neutrino Telescope on  AT2018cow~\citep{2018ATel11785....1B} already allows to exclude a region of the FBOT parameter space, otherwise compatible with electromagnetic observations. 
On the contrary, the existence of a fast outflow ($v_{\rm{ej}} \gtrsim 0.1$~c) interacting with the CSM is supported by electromagnetic observations. The results concerning the neutrino signal from CSM interaction are therefore robust and valid for any viewing angle of the observer, being the emission  isotropic in good approximation.

We find that the neutrino emission from LFBOTs does not account for the bulk of the diffuse neutrino background observed by IceCube. Nevertheless, the neutrino fluence from a single LFBOT is especially large in the choked jet scenario,  if the jet should be observed on axis, and  is comparable to the sensitivity of the IceCube Neutrino Observatory and IceCube-Gen2, while it is below the IceCube sensitivity for the CSM interaction cases.

By taking into account the Eddington bias on the observation of cosmic neutrinos, we conclude that the two track-like events observed by IceCube in coincidence with  AT2018cow may have been of astrophysical origin (similar conclusions would hold for CSS161010). In the light of these findings, a search for neutrino events  in coincidence with the other known LFBOTs should be carried out. 

In conclusions, the detection of neutrinos from LFBOTs with existing and upcoming neutrino telescopes will be crucial to probe the mechanism powering FBOTs. The choked jet and CSM interaction would generate very different neutrino signals: the former is direction dependent and peaks around $E_\nu \simeq 10^5$~GeV, the latter is quasi-isotropic and approximately flat up to $E_\nu \simeq 10^7$--$10^8$~GeV for our fiducial parameters. Current neutrino telescopes may not be able to clearly differentiate the signals from the choked jet and CSM interaction scenarios. Nevertheless, CSM interaction can produce neutrinos in the high energy  tail of the spectrum. E.g.~the  detection of neutrinos with energies of $\mathcal{O}(100)$~PeV may hint towards the CSM interaction origin; on the other hand, if a choked jet is harbored in LFBOTs and the jet is observed on-axis, a large number of neutrinos with $\mathcal{O}(100)$~TeV energy is expected to be detected at IceCube and IceCube-Gen2.
As the number of detected LFBOTs increases, neutrino searches have the potential to provide complementary information on the physics of these  emergent transient class and their rate. 

\section*{Acknowledgments}

We thank Erik Blaufuss for useful discussions as well as Ore Gottlieb and Brian Metzger for insightful comments on the manuscript. This project has received funding from the  Villum Foundation (Project No.~37358), the Carlsberg Foundation (CF18-0183), the Deutsche Forschungsgemeinschaft through Sonderforschungsbereich
SFB~1258 ``Neutrinos and Dark Matter in Astro- and Particle Physics'' (NDM), and  the National Science Foundation under Award Nos.~AST-1909796 and AST-1944985.

\appendix
\section{Proton and meson cooling times}
\label{sec:A}

For the choked jet case the acceleration time scale of protons is
\begin{equation}
t^{\prime -1}_{\rm{acc}} = \frac{c e B^{\prime}}{\xi E^{\prime}_p} \; ,
\end{equation}
where $e = \sqrt{\hbar \alpha c}$ is the electric charge with $\alpha = 1/137$ being the fine structure constant and $\hbar$ is the reduced Planck constant.  $\xi$ defines the number of gyroradii needed for accelerating protons, and we assume~$\xi = 10$~\citep{Gao:2012ay}. Finally, $B^\prime$ is the magnetic field generated at the internal shock, see main text.

For CSM interaction, the acceleration timescale  is obtained in the Bohm limit~\citep{protheroe_clay_2004}
 \begin{equation}
 t^{\prime -1}_{\rm{acc}} \simeq \frac{3 e B^\prime v_{\rm{sh}}^{2}}{20 \gamma_p m_{p} c^3 } \; ,
 \end{equation} 
where $B^\prime \equiv \tilde{B}$ is the magnetic field in the shocked interacting shell, see main text. 

Protons accelerated at the shocks undergo several energy loss processes. The total cooling time is 
\begin{equation}
t^{^\prime -1}_{p, \rm{cool}} = t^{^\prime -1}_{\rm{ad}} + t^{^\prime -1}_{p, \rm{sync}} +t^{^\prime -1}_{p \rm{\gamma}} + t^{^\prime -1}_{pp} + t^{^\prime -1}_{p, \rm{BH}} + t^{^\prime -1}_{p, \rm{IC}}\ ,
\label{eq:total_cooling}
\end{equation}
where $t^{^\prime -1}_{\rm{ad}}$, $t^{^\prime -1}_{p, \rm{sync}}$, $t^{^\prime -1}_{p \gamma}$, $t^{^\prime -1}_{{pp}}$, $t^{^\prime -1}_{p, \rm{BH}}$, $t^{^\prime -1}_{p, \rm{IC}}$ are the adiabatic, synchrotron, photo-hadronic ($p \gamma$), hadronic ($pp$), Bethe-Heitler (BH, $p \gamma \rightarrow p e^+ e^-$) and inverse Compton (IC) cooling timescales, respectively. These are defined as follows~\citep{dermer_book, Gao:2012ay, Razzaque:2005bh}: 
\begin{eqnarray}
 t^{\prime -1}_{\rm{ad}} &=& \frac{v}{R}\ , \label{eq:adiabatic_time} \\
 t^{\prime -1}_{p, \rm{sync}} &=& \frac{4 \sigma_T m_e^2 E^\prime_p B^{\prime 2}}{3 m_p^4 c^3 8 \pi}\ , \\
 t^{\prime -1}_{p \gamma} &=& \frac{c}{2 \gamma^{\prime 2}_p} \int_{E_{\rm{th}}}^\infty dE^\prime_\gamma \frac{n^\prime_{\gamma}(E^\prime_\gamma)}{E^{\prime 2}_\gamma} \int_{E_{\rm{th}}}^{2 \gamma^\prime_p E^\prime_\gamma} dE_r E_r \sigma_{p \gamma}(E_r) K_{p \gamma}(E_r)\ ,  \\
 t^{\prime -1}_{{pp}} &=& c n^\prime_{p} \sigma_{pp} K_{pp}\ ,  \\
 t^{\prime -1}_{p, \rm{BH}} &=& \frac{7 m_e \alpha \sigma_T c}{9 \sqrt{2} \pi m_p \gamma^{\prime 2}_p} \int_{\gamma_p^{\prime -1}}^{\frac{E^\prime_{\gamma, \rm{max}}}{m_e c^2}} d\epsilon^\prime \frac{n^\prime_{\gamma} (\epsilon^\prime)}{\epsilon^{^\prime 2}} \biggl\{ (2 \gamma^\prime_p \epsilon^\prime)^{3/2} \biggl[\ln(\gamma^\prime_p \epsilon^\prime) -\frac{2}{3} \biggr]+ \frac{2^{5/2}}{3} \biggr\}\ ,  \\
 t^{\prime -1}_{p, \rm{IC}} &=& \frac{3 (m_e c^2)^2 \sigma_T c}{16 \gamma_p^{\prime 2}( \gamma^\prime_p-1) \beta^\prime_p} \int_{E^\prime_{\gamma, \rm{min}}}^{E^\prime_{\gamma, \rm{max}}} \frac{dE^\prime_\gamma}{E_\gamma^{^\prime 2}}F(E^\prime_\gamma, \gamma^\prime_p) n^\prime_{\gamma}(E^\prime_\gamma)\ , 
\end{eqnarray}
where $v= 2 c \Gamma$ for the choked jet and $v = v_{\rm{sh}}$ for CSM interactions, $\gamma_p = E^\prime_p/m_p c^2$, $\epsilon^\prime = E^\prime_\gamma /m_e c^2$, $E_{\rm{th}}=0.150$~GeV is the energy threshold for photo-pion production, and $\beta^\prime_p \approx 1$ for relativistic particles. The function $F(E^\prime_\gamma, \gamma^\prime_p)$ follows the definition provided in~\cite{PhysRev.137.B1306},  replacing $m_e \rightarrow m_p$. The cross sections for $p \gamma$ and $pp$ interactions, $\sigma_{p \gamma}$ and $\sigma_{pp}$, can be found in~\cite{ParticleDataGroup:2020ssz}.
The function $K_{p\gamma}(E_r)$ is the $p\gamma$ inelasticity, given by Eq.~9.9 of~\cite{dermer_book}:
\begin{equation}
K_{p\gamma}(E_r) = 
\begin{system}
0.2 \; \; \; \; \; \;  \; \; \;  E_{\rm{th}} < E_r < 1~\rm{GeV}\\
0.6 \; \; \; \; \; \;  \; \; \;  E_r > 1~\rm{GeV}
\end{system} \
\end{equation}
where $E_r = \gamma^\prime_p E^\prime_\gamma (1 - \beta^\prime_p \cos\theta^\prime)$ is the relative energy between a proton with Lorentz factor $\gamma^\prime_p$ and a photon with energy $E^\prime_\gamma$, whose directions form an angle $\theta^\prime$ in the comoving frame of the interaction region. The comoving proton density is $n^\prime_{p} = 4 \tilde{L}_{j} /(4 \pi R_{\rm{IS}}^2 c m_p c^3 \theta_j^2)$ for the choked jet, and $n^\prime_{p} = \tilde{n}_p = 4 n_{p, \rm{CSM}} m_p c^2$ for CSM interaction. The inelasticity of  $pp$ interactions is $K_{pp} = 0.5$ and $n^\prime_{\gamma}(E^\prime_\gamma)$ is the photon target for accelerated protons.
\begin{figure*}[b]
\centering
\includegraphics[width=0.48\textwidth]{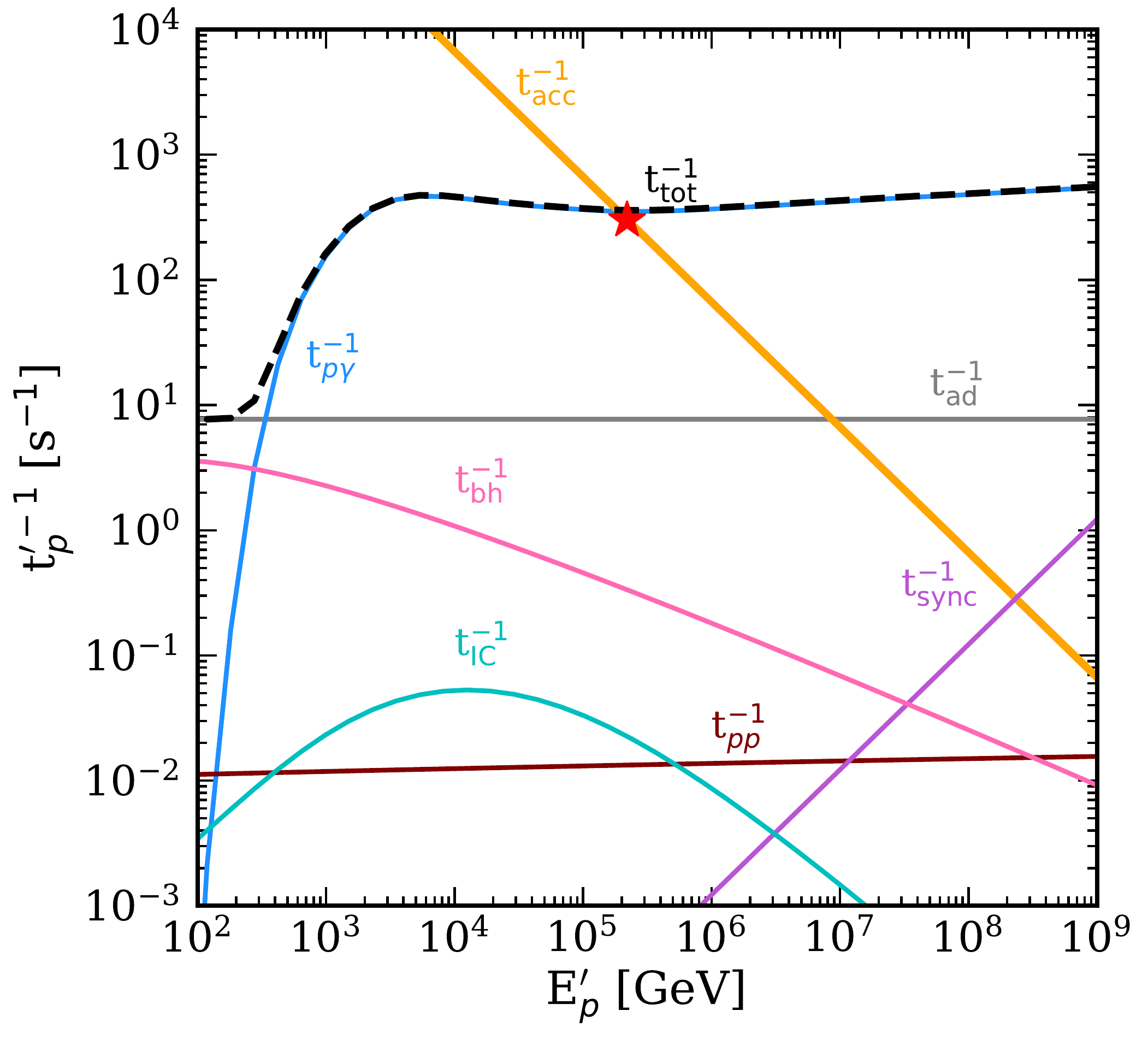}
\includegraphics[width=0.48\textwidth]{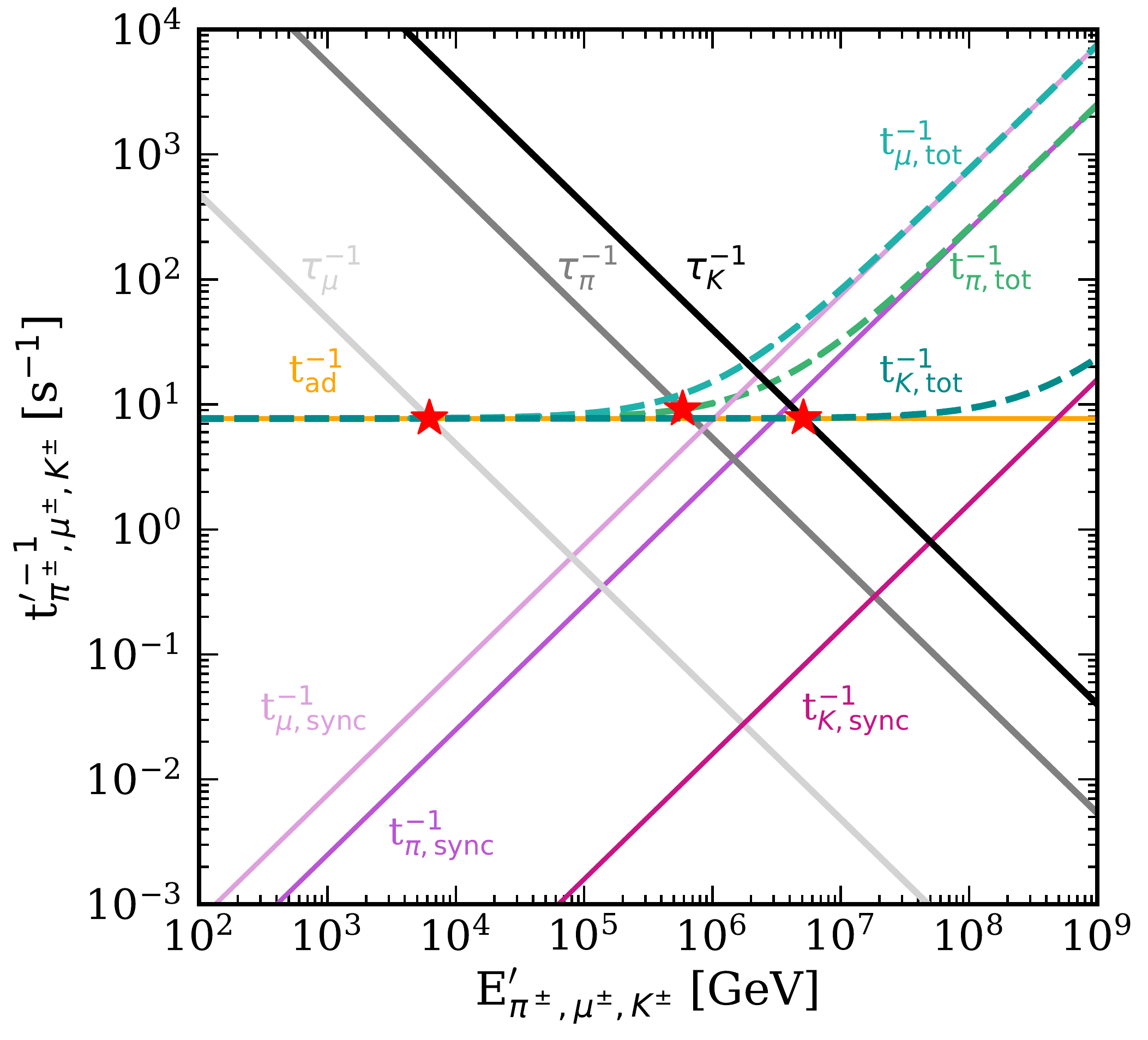}
\caption{Cooling times of protons accelerated (left panel) and   charged mesons (right panel) in the internal shock scenario as functions of the particle energy. Results are shown for $\tilde{L}_{\rm{iso}} = 5 \times10^{49}$~erg s$^{-1}$, $\tilde{t}_j = 20$~s and $\Gamma = 100$. The total cooling time is plotted with a dashed black line. For protons, $p \gamma$ interactions are the most efficient energy loss mechanism and they define the maximum energy of accelerated protons, marked with a red star. For mesons, adiabatic losses are the only relevant energy loss mechanism. The maximum energy that mesons can achieve before decaying is marked by  a red star.}
\label{fig:coolingchoked}
\end{figure*}
\begin{figure}
\centering
\includegraphics[width=0.48\textwidth]{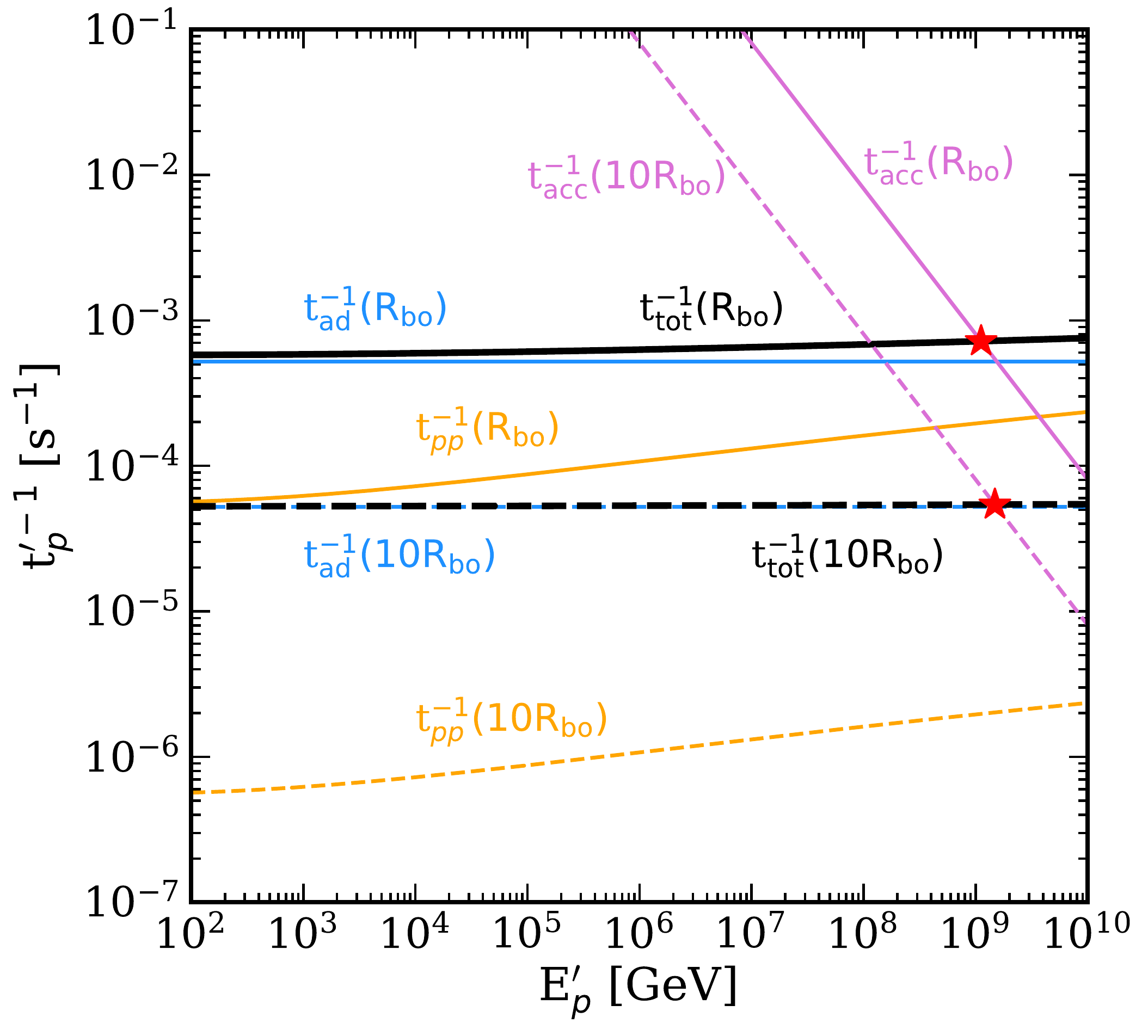}\caption{Cooling times of protons accelerated at the forward shock between the fast ejecta and the CSM as functions of the proton energy. We show the results at $R_{\rm{bo}}$ and $10 \times R_{\rm{bo}}$ for  $v_{\rm{ej }} = 0.55 c$, $M_{\rm{ej}} =3.7  \times 10^{-2} M_\odot$ and $M_{\rm{CSM}} = 10^{-1} M_\odot$. Adiabatic cooling is the most important energy loss mechanism, while $pp$ interactions are more competitive at the beginning of the evolution of the ejecta, but they rapidly drop. The red star marks the maximum energy that protons can reach in the shocked plasma shell. }
\label{fig:coolingpp}
\end{figure}

At the internal shock, secondary charged mesons undergo energy losses before decaying; in turn, affecting the neutrino spectrum. In Fig.~\ref{fig:coolingchoked}, we show an example  obtained for $\tilde{L}_{j} =  2 \times10^{47}$~erg s$^{-1}$, $\tilde{t}_j = 20$~s and $\Gamma = 100$. We note that, in the choked jet, $p \gamma$ interactions are the main energy loss channel for protons, while secondaries mainly cool through adiabatic losses. Kaons cool at energy much higher than the maximum proton energy, therefore their cooling does not affect the neutrino spectrum~\citep{He:2012tq, Asano:2006zzb, Petropoulou:2014lja, Tamborra:2015qza}. 

As for CSM interaction, the only relevant cooling processes for protons are  hadronic cooling ($pp$ interactions) and adiabatic cooling. The photons produced at the external shock between the ejecta and the CSM have energies in the radio band, i.e.~at  low energies. Therefore interactions between protons and photons are negligible, consistently with~\cite{Murase:2010cu,Fang:2020bkm}. For CSM interaction, ${t}^{\prime -1}_{\rm{cool}} = {t}^{\prime -1}_{pp} + {t}^{\prime -1}_{\rm{ad}} $ (note that since shocks are non-relativistic, there is no difference between the comoving frame of the shock and the CO frame for CSM interaction). The proton cooling times are shown  at $R_{\rm{bo}}$ and $10 R_{\rm{bo}}$ in Fig.~\ref{fig:coolingpp} for $v_{\rm{ej }} = 0.55 c$, $M_{\rm{ej}} = 3.7 \times 10^{-2} M_\odot$ and $M_{\rm{CSM}} = 10^{-1} M_\odot$. We note that the $pp$ interactions are more efficient at earlier times, though they are less important than adiabatic losses throughout the ejecta evolution, as expected  because of the low densities of the CSM.

 \bibliography{references}{}
\bibliographystyle{aasjournal}

\end{document}